\documentclass[12pt,thmsa,fleqn]{article}
\usepackage[left=1in,right=1in,top=1in,bottom=1in]{geometry}
\newcommand{\qed}{\nobreak \ifvmode \relax \else
	\ifdim\lastskip<1.5em \hskip-\lastskip
	\hskip1.5em plus0em minus0.5em \fi \nobreak
	\vrule height0.5em width0.7em depth0.05em\fi}
\usepackage{graphicx}
\graphicspath{{Figures/}}
\usepackage{amsmath,bm}
\usepackage{float}
\usepackage{epsfig}
\usepackage{times}
\usepackage{setspace}
\usepackage{titlesec}
\usepackage{abstract}
\usepackage[numbers]{natbib}

\titleformat{\section}
{\normalfont\fontsize{14}{15}\bfseries}{\thesection}{1em}{}
\titleformat{\subsection}
{\normalfont\fontsize{13}{15}\bfseries}{\thesubsection}{1em}{}
\titleformat{\subsubsection}
{\normalfont\fontsize{12}{15}\bfseries}{\thesubsubsection}{1em}{}
\usepackage{multicol}  
\usepackage[mathscr]{eucal}
\usepackage{amssymb}
\usepackage{wrapfig}
\usepackage{physics}
\makeatletter
\def\Let@{\def\\{\notag\math@cr}}
\makeatother

\makeatletter

\newcommand{\Matrix}[1]
{\begin{pmatrix}
		\Matrix@r #1;\@bye;\Matrix@r
\end{pmatrix}}

\def\Matrix@r #1;{\@bye #1\Matrix@z\@bye\Matrix@s #1,\@bye, }%
\def\Matrix@s #1,{#1\Matrix@t }%
\def\Matrix@t #1,{\@bye #1\Matrix@y\@bye\@firstofone {&#1}\Matrix@t}%
\def\Matrix@y #1\Matrix@t{\\ \Matrix@r }%
\def\Matrix@z #1\Matrix@r {}
\def\@bye  #1\@bye   {}

\makeatother

\usepackage{array}

\usepackage[
colorlinks=true,
linkcolor=blue,
urlcolor=blue,
citecolor=blue]{hyperref}

\usepackage{caption,subcaption}

\begin{document}

\begin{center}
{\Large  The concealment of  accelerated information is possible}
\end{center}

\begin{center}

A.G. Abdelwahab$^{1}$, Nasser Metwally $^{2,3}$, M.H. Mahran$^{1}$, A-S F Obada$^{4}$ \\
 $^{1}$\textit{Department of Basic Science, Faculty of Computers and Informatics, Suez Canal University, Ismailia 41522, Egypt.}\\
$^{2}$\textit{ Department of Mathematics., College of Science, University of Bahrain, 320038, Bahrain}\\
$^{3}$\textit{Department of Mathematics, Faculty of science, Aswan University,  Aswan, Sahari 81528, Egypt}\\
$^{4}$\textit{ Math. Dept., Faculty of Science, Al-Azhar University, Nasr City 11884, Cairo, Egypt}\\
\end{center}

\begin{abstract}

	The possibility of masking an accelerated two-qubit system by using  a minimum number of qubits is discussed. It is shown that, the information  may be  masked in either entangled local states or product non-local  separable states.  We examine that each partition of  these states satisfies the masking conditions.  Due to the presence of non-local separable partition, one may consider that it is a type of quantum data hiding scheme.  The local /non-local information encoded in the masked  entangled  state is robust against the decoherence of the acceleration process. The possibility of estimating the acceleration parameter via the entangled/separable masked state increases as the initial entanglement value increases. The efficiency of  the masking process is examined by quantifying the fidelity of the accelerated state and its subsystems. It is shown that, the fidelity of the masked state is maximum at small initial acceleration, while the minimum fidelity is  more than $96\%$.

\end{abstract}


\section{Introduction}
\label{intro}
Saving information represents one of the most important issues that disturbs   their owners. Therefore, practically  storing them or sending them safely a crucial step to start any project.  However, there are many protocols that have been introduced for this purpose. To secure communication , quantum cryptography \cite{Gisin2002} is one of the most techniques that has been developed and examined from  different points of view. On the other hand, the concealment of information is an essential technique that protect handling information securely \cite{Fink2017}. While quantum coding \cite{Schumacher1995} is a powerful method to encode information that can be used to  store and send information safely.

We may mention that quantum masking is another tool that could be used to develop  the recent techniques  of keeping information security. Recently, there are some attempts  introduced, to understand the masking process on entangled systems.  For example, Modi et al.\cite{Modi2018} defined  some conditions that must be satisfied  by any masker of quantum information.  Ghosh et al. \cite{Ghosh2019}  verified the possibility of masking quantum information under some restricted conditions. The optimal number of parties that can be used to mask any arbitrary state  is discussed in Ref. \cite{Li2019}. The problem of information masking through nonzero linear operators is discussed by \cite{Liang2020,Liang2019}. The quantum masker has been used to discuss the unconditionally secure qubit commitment \cite{Lie2019}.

Recently, accelerated quantum systems have attracted the attention of many researchers from different aspects. There are some efforts to quantify the  amount of entanglement that contained on these accelerated systems\cite{Fuentes-Schuller2005,Alsing2006,Moradi2009,Landulfo2009,Martin-Martinez2010a, metwally2016}.  Metwally  discussed the possibility of using them to perform some quantum information tasks as teleportation\cite{metwally2013,Hamdoun2020} dense coding \cite {metwally2014}. The quantum security over accelerated quantum channel is discussed by \cite{Bradler2008}.

 Therefore, in this contribution we are motivated to  discuss the possibility of applying the simplest masking protocol on the accelerated systems. For this purpose,  it is  assumed that an accelerated two-qubit system is initially prepared in the $X$-state. In this treatment, we assume that either both or  a single qubit is accelerated.  The masking process is applied only on the accelerated subsystem(s). We examine the masking conditions on all the possible partitions and we discuss  whether they contain classical or quantum correlations. The local/non-local masked information as well as the masked  quantum/ classical Fisher information are quantified. The fidelity of the masked accelerated state and its marginal subsystems is examined.

The paper is organized as follows. In Sec. \ref{system}, we review the accelerated initial state, where we quantify the initial entanglement by using the negativity. The mathematical forms of the Fisher information  and the local/ non-local information are introduced. Moreover, all the three quantities are given for the accelerated system. In Sec. \ref{masking}, the masking process is applied on the accelerated system, where we assume that both qubits are accelerated and masked.  Additionally,  we show that all the masked states satisfy the masking conditions. We investigate numerically, the behavior of the masked Fisher information, local and non-local information for some initial states settings. Sec. \ref{Alice-accelerated} is devoted to discuss the masking process when only one qubit is accelerated. In Sec. \ref{fidelity}, we discuss the accuracy of the masking process and its marginal parts by calculating  the fidelity of the accelerated masked state and its subsystems.  Finally, we discuss our results and conclude the paper in Sec. \ref{conclusion}.

\section{The proposed system}
\label{system}
Let us assume that, the users Alice and Bob   share a two qubit state of the $X$- state type.  In the computational basis \cite{Yu2007a}, this state may be written as,

\begin{align}
\rho&=\mathcal{A}_{11}\ket{00}\bra{00}+\mathcal{A}_{14}\ket{00}\bra{11}+\mathcal{A}_{22}\ket{01}\bra{01}
+\mathcal{A}_{23}\ket{01}\bra{10}
\\& \quad +\mathcal{A}_{32}\ket{10}\bra{01}+\mathcal{A}_{33}\ket{10}\bra{10}
 +\mathcal{A}_{41}\ket{11}\bra{00} +\mathcal{A}_{44}\ket{11}\bra{11} ,
\label{GWSeq:rhoprev}
\end{align}
where,
\begin{align}
&\mathcal{A}_{11}=\mathcal{A}_{44}=\frac{1+z}{4},~\quad\mathcal{A}_{22}=\mathcal{A}_{33}=\frac{1-z}{4},~\quad
\mathcal{A}_{23}=\mathcal{A}_{32}=\frac{x+y}{4}, \\
&\mathcal{A}_{14}=\mathcal{A}_{41}=\frac{x-y}{4},~\quad |\eta|\leq 1,\eta=x,y,z
\label{GWSArelation}
\end{align}

 To examine the masking process on the accelerated two systems, we discuss the behavior of the initial state (\ref{GWSeq:rhoprev}) in the
non-inertial frames.  In this context, it is important to remind ourselves by  Unruh effect on
Minkowski's particles. It is well known that the  Minkowsik
coordinates $(t,\gamma)$ can be used to investigate the dynamics of
these particles in the  inertial frames. However the  Rindler
coordinates $(\zeta, \chi)$ may be used to describe the system in the non-inertial frame. These coordinates are connected by the following relations,
\begin{equation}\label{trans}
\zeta=r~tanh\left(\frac{t}{\gamma}\right), \quad \chi=\sqrt{\gamma^2-t^2},
\end{equation}
where  $-\infty<\zeta<\infty$, $-\infty<\chi<\infty$  and $r$ is
the acceleration of the moving particle. The transformation
(\ref{trans}) defines two regions in Rindler's spaces: the first
region $I $ for $|t|<\gamma$  and the second region $II$ for $\gamma<-|t|$
\cite{Martin-Martinez2011}.
In terms of
Rindler's modes, the  Minkowski vacuum $\ket{0_k}_M$ and the one
particle
 state $\ket{1_k}_M$
 take the form,
\begin{eqnarray}\label{vacuum1}
\ket{0_k}_M&=&\cos r\ket{0_k}_I\ket{0_{-k}}_{II}+ \sin
r\ket{1_k}_I\ket{1_{-k}}_{II}, \nonumber\\
\ket{1_k}_M&=&\ket{1_k}_I\ket{0_k}_{II}.
\end{eqnarray}
where, $c_i=\cos r_i,~ s_i=sin r_i,$ with
$tan r_i=e^{-\pi\omega_i \frac{c}{a_i}}$, $a_i$ is the
acceleration, $\omega_i$ is the frequency of the traveling
qubits, $c$ is the speed of light and $i=A,R$ stands for Alice and
Bob respectively. In this investigation, it is assumed that, either Alice/
 Bob or both of them are accelerated. Therefore, a uniformly accelerated observer lying in
one wedge of space time is causally disconnected from the other.
Since, we are interested on the connected accelerated state between Alice and Bob, $\rho_{acc}$ is in  region $I$,
one has  to trace out the state of  Anti-Alice and Anti-Bob in the region $II$.
Now, by using the acceleration process, then the accelerated state between Alice and Bob in the first region $I$ is given by,
\begin{align}
\rho_{\text{acc}}&= \mathcal{B}_{11} \ket{00}\bra{00}+\mathcal{B}_{14} \ket{00}\bra{11}+\mathcal{B}_{22} \ket{01}\bra{01} + \mathcal{B}_{23} \ket{01}\bra{10}
 \\& \quad+\mathcal{B}_{32}\ket{10}\bra{01} +\mathcal{B}_{33} \ket{10}\bra{10}
 +\mathcal{B}_{41}\ket{11}\bra{00}+\mathcal{B}_{44} \ket{11}\bra{11},\\
\label{GWSrhoaccelerated}
\end{align}
where
\begin{align}
&\mathcal{B}_{11} = c^4 \mathcal{A}_{11} ,  \quad
\mathcal{B}_{14} = c^2 \mathcal{A}_{14} ,
\mathcal{B}_{22} = c^2 \left(s^2 \mathcal{A}_{11} + \mathcal{A}_{22} \right),  \quad \mathcal{B}_{23} = c^2 \mathcal{A}_{23}  ,  \quad
\mathcal{B}_{32} = c^2 \mathcal{A}_{32}  ,  \quad\\&
\mathcal{B}_{33} = c^2 \left(s^2 \mathcal{A}_{11} + \mathcal{A}_{33} \right) ,  \quad
\mathcal{B}_{41} = c^2 \mathcal{A}_{41}  ,  \quad
\mathcal{B}_{44} = s^4 \mathcal{A}_{11} + s^2 \mathcal{A}_{22} + s^2 \mathcal{A}_{33} +\mathcal{A}_{44}.
\end{align}
Since the amount of entanglement that contained in the accelerated state between the users is an essential quantity to perform many quantum tasks, it is important to quantify it. For this aim, we consider the negativity as a common measure of entanglement.
However, for any two qubit system  $\rho_{AB}$, the negativity is defined as	\cite{Peres1996,Horodecki1996}
\begin{equation}
	\mathcal{N}(\rho_{AB}) = \sum_{i} \left| \lambda_{i}\right| -1
\end{equation}
where $\lambda_{i}$ are the eigenvalues of the partial transpose of $\rho_{AB}$ with respect to the second subsystem $B$
\begin{figure}
	\centering
	{\includegraphics[width=3in,height=3in]{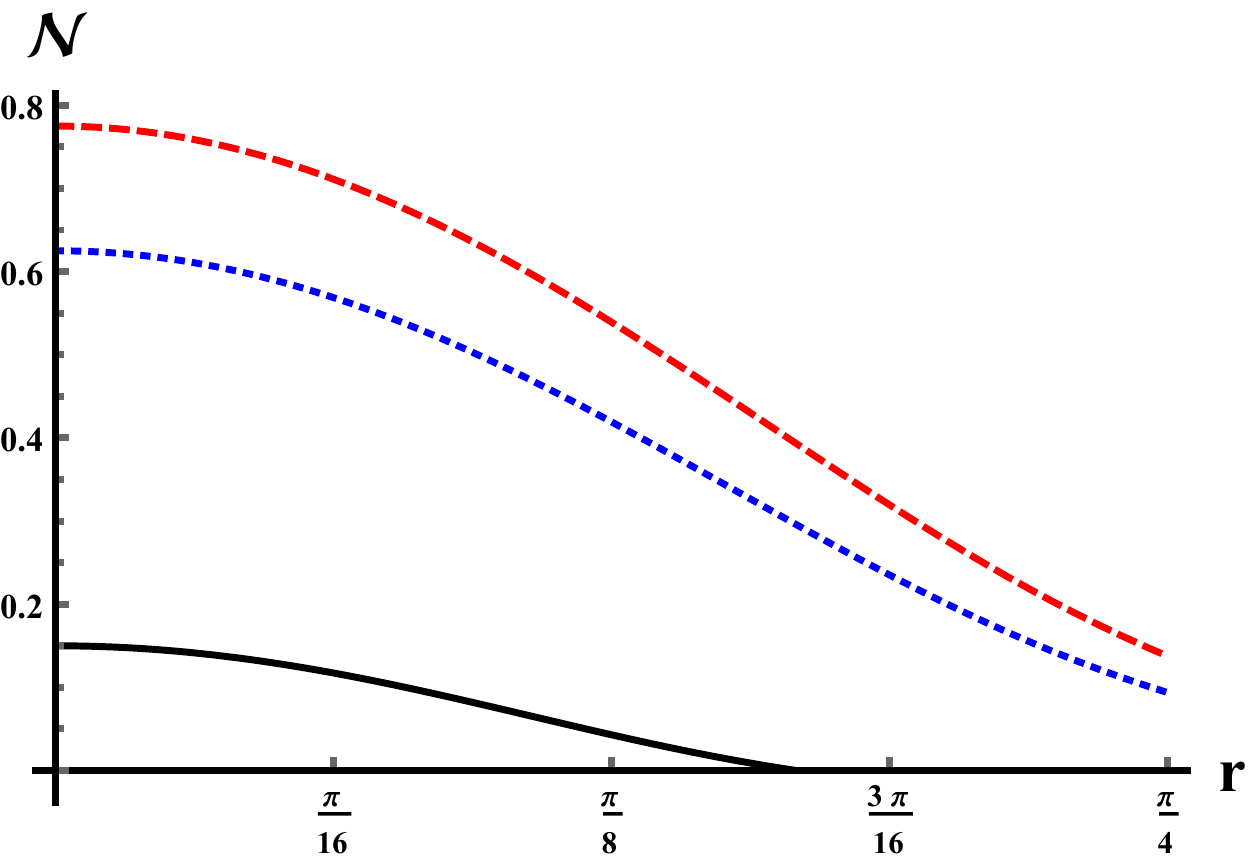}}
	\caption{ The negativity for the state \eqref{GWSrhoaccelerated} initially prepared with $x = y = z = -0.85$ for the dashed curve, the Werner state with $x = y = z = -0.75$ for the dotted curve and the generalized Werner state with $x = - 0.6$, $y = - 0.2$ and $z = - 0.5$ for the solid curve.}
	\label{FGWS9fig:negativityacc}
\end{figure}
The behavior of negativity  $\mathcal{N}(\rho_{AB})$ of the accelerated state \eqref{GWSrhoaccelerated}  is shown in \figurename{(\ref{FGWS9fig:negativityacc})}, where different initial state settings are considered. It is clear that the amount of the survival entanglement depends on the initial entanglement, where for the non-accelerated state, ie. $r=0$, the negativity is maximum. However as the the acceleration increases, the negativity decays, meaning: the amount of the survival entanglement decreases. The decay increases if the initial entanglement is small.
\subsection{Initial accelerated Information}
\label{initialsys}
In this section, we review some types of information that can be distilled from the accelerated state  \eqref{GWSrhoaccelerated}.
Among these types we mention the quantum Fisher information, local and non-local information.
\subsubsection*{Quantum Fisher Information:}

The quantum Fisher information (QFI) with respect to any parameter describes the sensitivity of the state according to changes of this parameter. Let the parameter $\theta$ is coded in the state  $\rho_{\theta}$, then the QFI $\mathcal{F}_{\theta}$ with respect to the parameter $\theta$ is defined as \cite{Yao2014}
\begin{align}
\mathcal{F}_{I}^{\theta}&=\sum_{i=1}^{n} \frac{1}{ \lambda_{i}} \left( \frac{\partial \lambda_{i}}{\partial \theta}\right)^{2}
\\& \quad +4 \sum_{i=1}^{n} \lambda_{i} \left( \bra{\frac{\partial V_{i}}{\partial \theta}}\ket{ \frac{\partial V_{i}}{\partial \theta}}- \left|\bra{ V_{i}}\ket{\frac{\partial V_{i}}{\partial \theta }} \right|^{2} \right)
\\& \quad - 8 \sum_{i\neq j}^{n}  \frac{ \lambda_{i} \lambda_{j} }{ \lambda_{i} + \lambda_{j} } \left| \bra{V_{i}}\ket{\frac{\partial V_{j}}{\partial \theta}}\right|^{2},
\end{align}
where $\lambda_{i}$ and $\ket{V_{i}}$ are the eigenvalues and  the corresponding eigenvectors of $\rho_{\theta}$. It is clear that, the accelerated state \eqref{GWSrhoaccelerated} depends on the external parameter, $r$ and the initial state parameters settings. Therefore to estimate these parameters one has to evaluate the eigenvalues and the eigenvectors in the normal computational basis, $00,01,10,11$ as,

\begin{align}
&\lambda_{1,2}=-\frac{1}{8} \cos^2 r (-3+z  \pm 2(x+ y) +(1+z) \cos (2 r)), \\
&\lambda_{3,4}=\frac{1}{32} \left(3 z+11 + 4 (z-1) \cos (2 r) + (z+1) \cos (4 r)-2\kappa (x,y)\right), \\
&\ket{V_{1}}=\left(0,-\frac{1}{\sqrt{2}},\frac{1}{\sqrt{2}},0 \right), \quad \ket{V_{2}}=\left(0,\frac{1}{\sqrt{2}},\frac{1}{\sqrt{2}},0 \right), \\
&\ket{V_{3}}=\frac{1}{\sqrt{1+\mu_{1}^{2}}} \left(\mu_{1},0,0,1 \right), \quad \ket{V_{4}}=\frac{1}{\sqrt{1+\mu_{2}^{2}}} \left(\mu_{2},0,0,1 \right) ,
\end{align}
where

\begin{align*}
\mu_{1,2} (r)&=\frac{\sec^2 r}{4 (x-y)} \left(-4+4 \cos (2 r)\mp \kappa (x,y)\right) .\\
\kappa (x,y)&=\sqrt{8 \cos (2 r) \left((x-y)^2-4\right)+2 (3+\cos (4 r)) \left((x-y)^2+4\right)}.
\end{align*}
\begin{figure}[t!]
	\begin{subfigure}[t]{0.5\textwidth}
		\centering
		\caption{}
		\includegraphics[width=3in,height=3in]{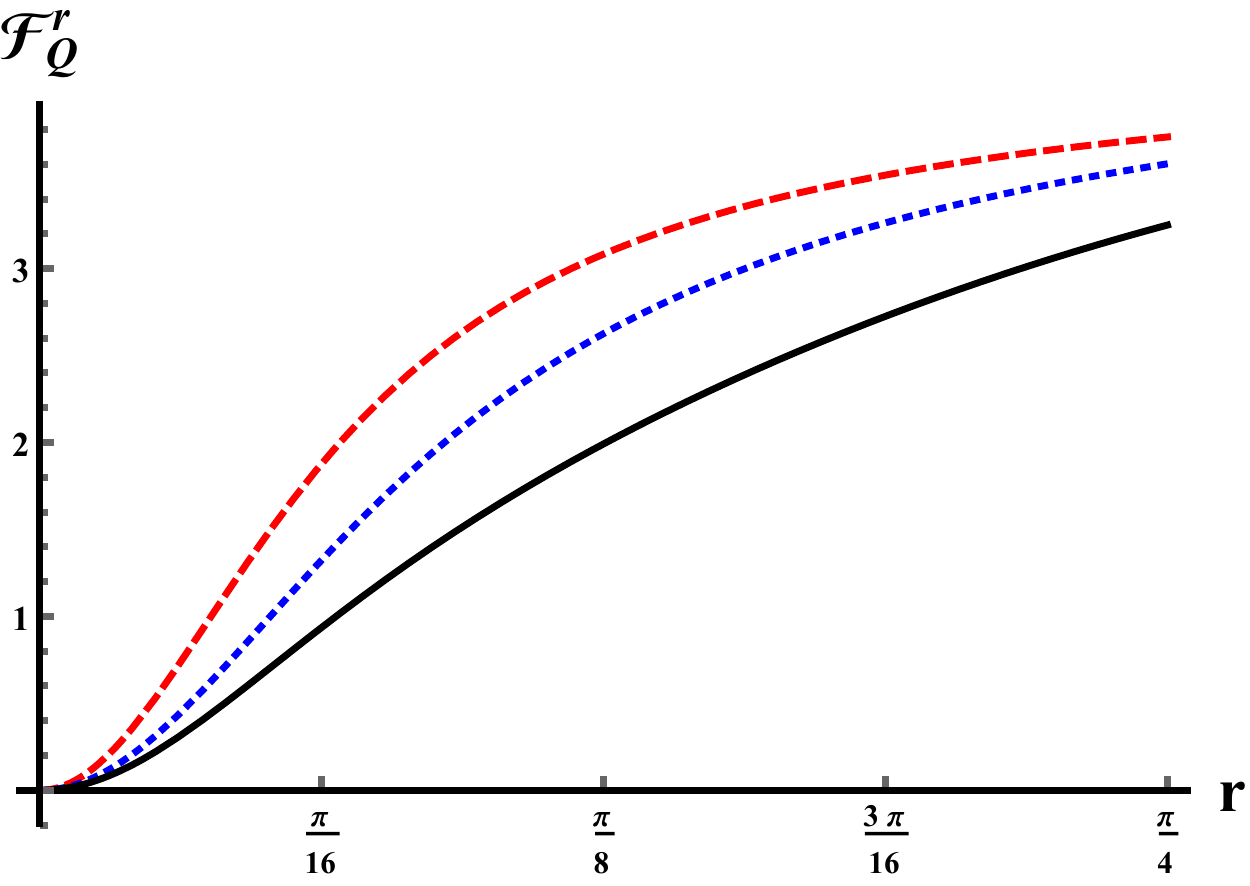}
		\label{FGWS9fig:QFbeforemaskingr}
	\end{subfigure}
~
	\begin{subfigure}[t]{0.5\textwidth}
		\centering
		\caption{}
		\includegraphics[width=3in,height=3in]{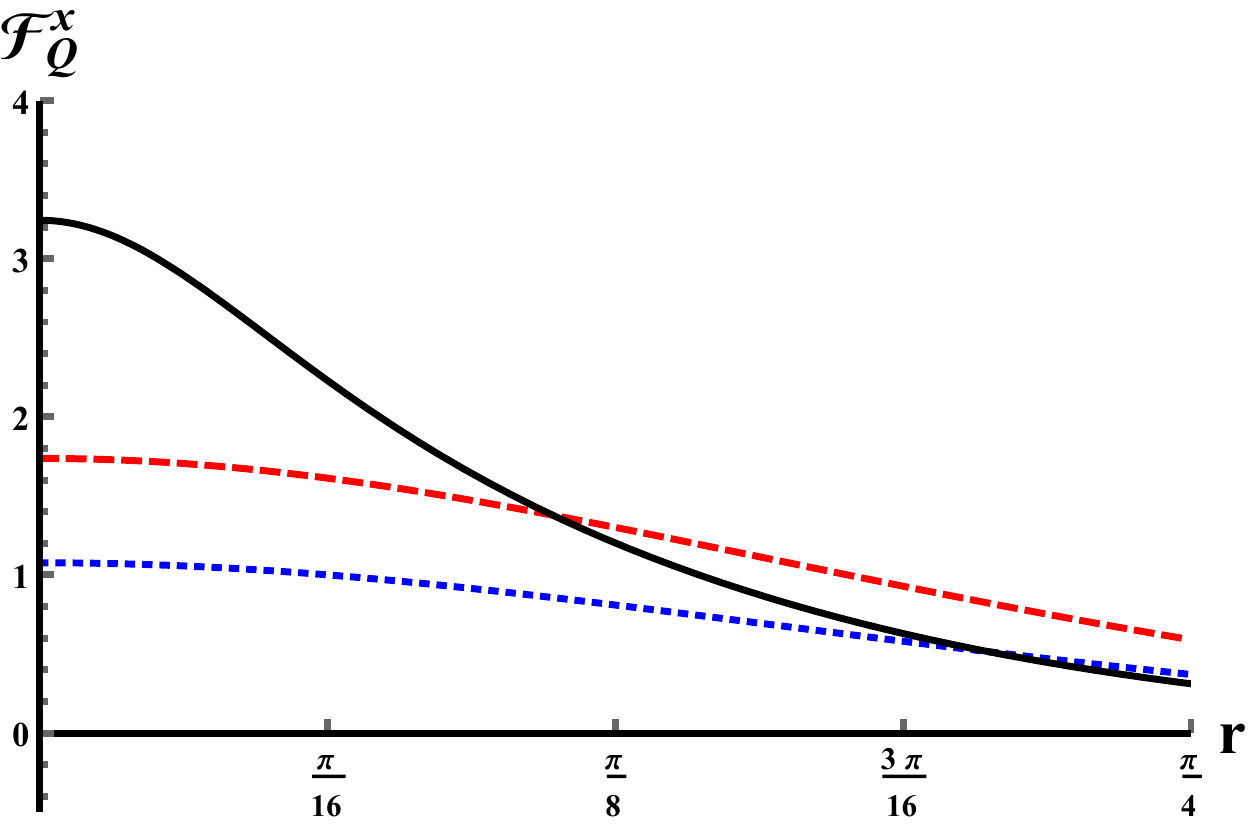}
		\label{FGWS9fig:QFbeforemaskingx}
	\end{subfigure}
	\caption{{ Estimating  (\subref{FGWS9fig:QFbeforemaskingr}) the acceleration parameter $r$, and (\subref{FGWS9fig:QFbeforemaskingx}) the $x$ parameter by using  \eqref{GWSrhoaccelerated}. It is assumed that the system is initially prepared in a Werner states  with $x = y = z = -0.85$ for the dashed curve, $x = y = z = -0.75$ for the dotted curve and the $X$- state with $x = - 0.6$, $y = - 0.2$ and $z = - 0.5$ for the solid curve.}}
	\label{FGWS9fig:QFbeforemasking}
\end{figure}
In \figurename(\ref{FGWS9fig:QFbeforemasking}), we estimate the acceleration parameter $r$ and estimate one parameter of the initial state settings.  For this estimation process, we evaluate the quantum Fisher information with respect to the parameters $r$ and $x$.  As it is displayed; from \figurename(\ref{FGWS9fig:QFbeforemaskingr}) the possibility of estimating the parameter $r$ increases as the acceleration increases. The increasing estimation degree of the parameter $r$, depends on the  initial entanglement of the accelerated system.  Therefore, the largest possibility of estimating the acceleration $r$ is depicted for an accelerated system initially prepared in a maximum entangled state. The behavior of the quantum Fisher information with respected to the parameter $x$, $\mathcal{F}_Q^x$ is displayed in \figurename(\ref{FGWS9fig:QFbeforemaskingx}). It is clear that the maximum values of $\mathcal{F}_Q^x$ is shown at $r=0$. As soon as the acceleration process  starts ,$\mathcal{F}_Q^x$  decreases gradually as the acceleration increases.
\subsubsection*{Local and Non-Local Information:}
The local information $\mathcal{I}_{loc}$ for a qubit  and the non-local  $\mathcal{I}_{nonloc}$ information between the two qubits are defined by \cite{Metwally2016a,Abd-Rabbou2019a}
\begin{equation}
	\mathcal{I}_{j}=\log_{2} D- \mathcal{S}(\rho_{i})
\end{equation}
where $j= loc$ and $nloc$ for the local and non-local information, respectively. $ \mathcal{S}$ is the von-Neumann entropy and $D$ is the dimension of the state $\rho_{i}, i=a(b), ab$.
\begin{figure}[t!]
	\begin{subfigure}[t]{0.5\textwidth}
		\centering
		\caption{}
		\includegraphics[width=3in,height=3in]{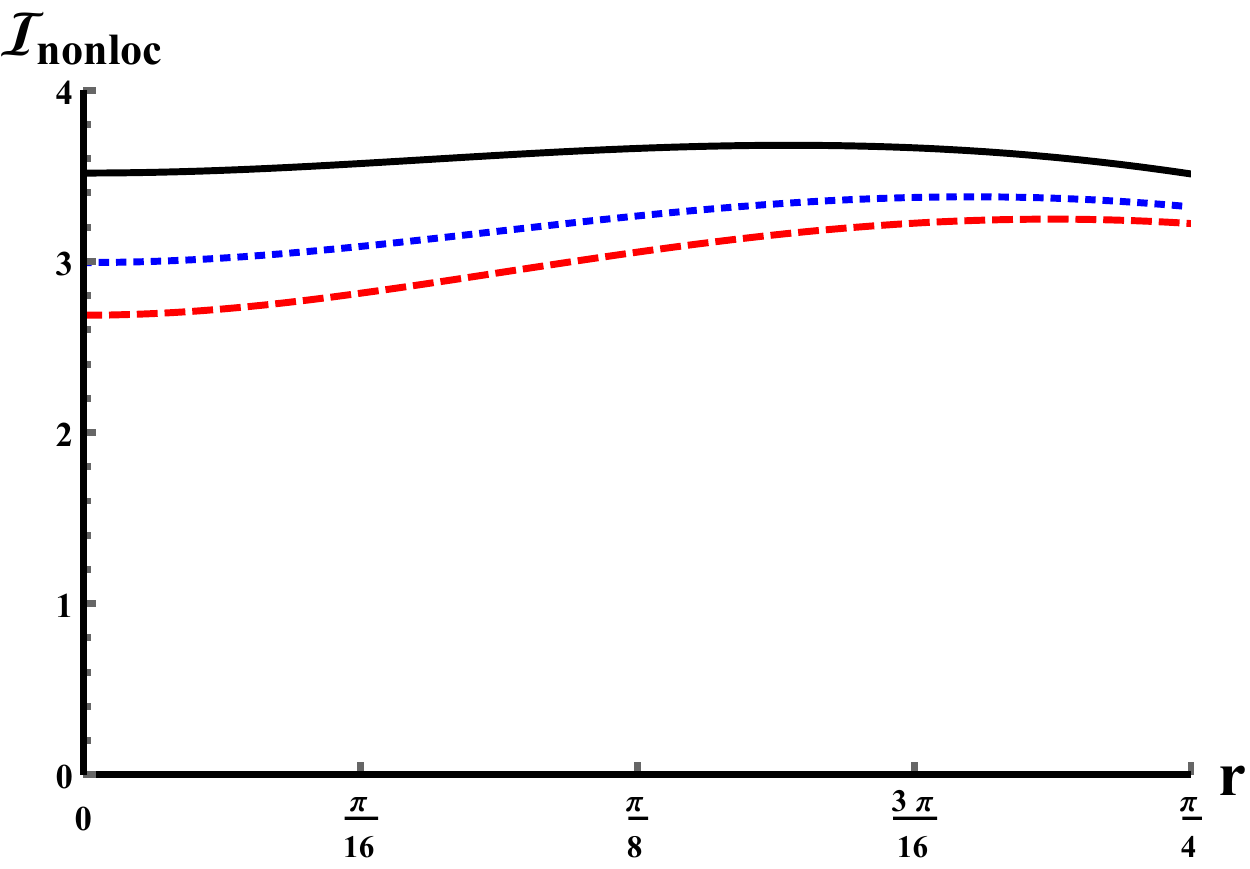}
		\label{FGWS9fig:nlocbefmask}
	\end{subfigure}
	~
	\begin{subfigure}[t]{0.5\textwidth}
		\centering
		\caption{}
		\includegraphics[width=3in,height=3in]{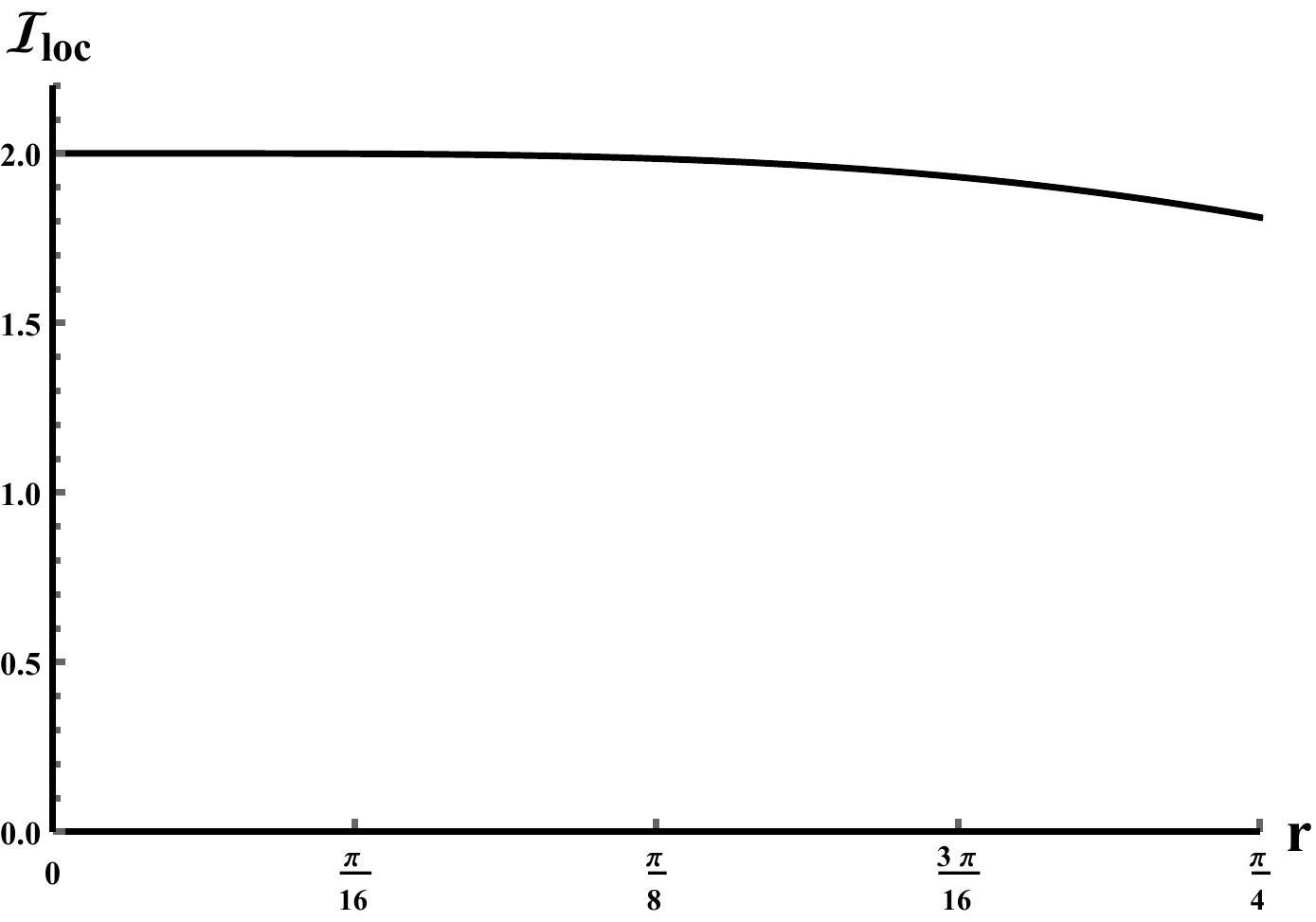}
		\label{FGWS9fig:locbemask}
	\end{subfigure}
	\caption{{ The local and non-local information for the state \eqref{GWSrhoaccelerated}, where the  system is initially prepared in  Werner state with $x = y=z=-0.85$, for the dashed curve,  $x = y=z=- 0.75$, for the dotted curve and the  $X$- state with $x = - 0.6$, $y = - 0.2$ and $z = - 0.5$ for the solid curve.}}
	\label{FGWS9fig:figureloc}
\end{figure}
The behavior of the local and non-local information of the accelerated state  \eqref{GWSrhoaccelerated} is displayed in Fig.(\ref{FGWS9fig:figureloc}). Since the two qubits are accelerated, then the local and non-local information are slightly affected by the acceleration. On the other hand, the less entangled state, has a high ability to resist the decoherence which is a result of the acceleration process. However, as one increases the initial entanglement, the robustness of the decoherence decreases.
\section{Masking  Process}
\label{masking}
The masking process of a two qubit-system is defined as: one says that an operator $\mu$ is a masker operator if it maps the state $\rho_{ab}$ into $\rho_a\otimes\rho_b$, where $\rho_{ab}\in \mathcal{H}_{ab}$ and $\rho_{a}\otimes\rho_{b}\in \mathcal{H}_a\otimes\mathcal{H}_b$.
 The marginal states are defined as: $\rho_a=Tr_b\{\rho_{ab}\}$ and $\rho_b=Tr_a\{\rho_{ab}\}$. This means that, the masked state is distributed into to two parties who has no information about the initial state.

In this section, we examine the possibility of using the most simple protocol to mask the accelerated system  \eqref{GWSrhoaccelerated}. This protocol suggests that each single information is masked by using two qubits \cite{Ghosh2019}, where each user can generate his/her  quantum circuit for the masking purpose. In this context, each $ \ket{0}\rightarrow \frac{\ket{00}+\ket{11}}{\sqrt{2}}$ and $\ket{1}\rightarrow \frac{\ket{00}-\ket{11}}{\sqrt{2}}$, which represents the minimum numbers of masker qubits.


\subsection{Both qubits are accelerated and masked}
\label{Both-accelereted-masked}
 Now, by applying this protocol on our suggested accelerated state \eqref{GWSrhoaccelerated}, one gets the total masked state as,

\begin{eqnarray}
\rho^{mask_{ab}}_{\text{acc}}&=&\frac{1}{4}\Bigl\{
R^{+} \left( \ket{0000} \bra{0000} + \ket{1111} \bra{1111} \right)+
R^{-} \left( \ket{0011}\bra{0011} + \ket{1100}\bra{1100} \right)
\nonumber\\
&+&U^{-} \left(  \ket{0000}\bra{1111} + \ket{1111}\bra{0000} \right)+
U^{+} \left(   \ket{0011}\bra{1100} + \ket{1100}\bra{0011}   \right)
\nonumber\\
&-&\sin ^2r\Bigl(\ket{0000}\bra{0011}  + \ket{0000}\bra{1100}  +  \ket{0011}\bra{0000} + \ket{1100}\bra{0000}
 \nonumber\\
 &+& \ket{1111}\bra{0011}+ \ket{1111}\bra{1100}+ \ket{0011}\bra{1111} + \ket{1100}\bra{1111}\Bigr)\Bigr\},
\end{eqnarray}

where $R^{\pm}=\left(1 + x \cos ^2r \right)$ and $U^{\pm}=\left( z \cos ^4r-y \cos ^2r+\sin ^4r \right)$.\\
In this context, we have to check whether this protocol masks the accelerated state. In other words, are the  masking conditions satisfied?. For this aim, one has to evaluate all the possible partitions of  the reduced density operators. Our calculations show that,

\begin{align}
\rho^{mask_{ab}}_{12}&=\rho^{mask_{ab}}_{34}\\
&=\frac{1}{2}\left( \ket{00} \bra{00} + \ket{11} \bra{11} \right) -\frac{1}{2}\sin^2 r \left(  \ket{11}\bra{00} + \ket{00}\bra{11} \right).
\label{FGWS9rho12}
\end{align}
and,
\begin{align}
	\rho^{mask_{ab}}_{13}&=\rho^{mask_{ab}}_{23}=\rho^{mask_{ab}}_{14}=\rho^{mask_{ab}}_{24} \\
	&=\kappa^+\left( \ket{00}\bra{00}+\ket{11}\bra{11}\right)
	 +\kappa^- \left(  \ket{01}\bra{01} + \ket{10}\bra{10} \right).
	\label{FGWS9rho13}
\end{align}
where $\kappa^{\pm}=\frac{1}{4}(1\pm x\cos^2r)$.

It is clear that, for all these partitions $\rho^{mask_{ab}}_1=\rho^{mask_{ab}}_2=\rho^{mask_{ab}}_3=\rho^{mask_{ab}}_4=\frac{1}{2}I_{2}$. This means that the masking conditions are satisfied, and consequently all the information are depicted only on the generated entangled states. Therefore, it is important to quantify the amount of entanglement generated in all the reduced density operators. The negativity of  the state  \eqref{FGWS9rho12}  is shown in \figurename{(\ref{FGWS9fig:negativity12})}. It is clear that, the negativity does not depend on the type of the initial state. It depends only on the accelerated parameter, where it  increases as $r$ increases.\\
The quantum Fisher information  with respect to the parameter can be evaluated explicitly by using the state  \eqref{FGWS9rho12} as,
\begin{equation}
\mathcal{F}_{Q}^{r}(r)=2 \sin ^2 r +\frac{\sin ^2 2r}{3-\cos 2 r}.
\end{equation}
The behavior of the quantum Fisher information $\mathcal{F}_{Q}^r$ is displayed in \figurename{(\ref{FGWS9fig:QFaftermaskingr12})}. It is clear that, $\mathcal{F}_{Q}^r$ increases as $r$ increases, where the maximum value of $\mathcal{F}_{Q}^r=4/3$ is obtained at $r=\pi/4$.
{\normalsize
\begin{figure}[t!]
	\begin{subfigure}[t]{0.5\textwidth}
		\centering
		\caption{}
		\includegraphics[width=3in,height=3in]{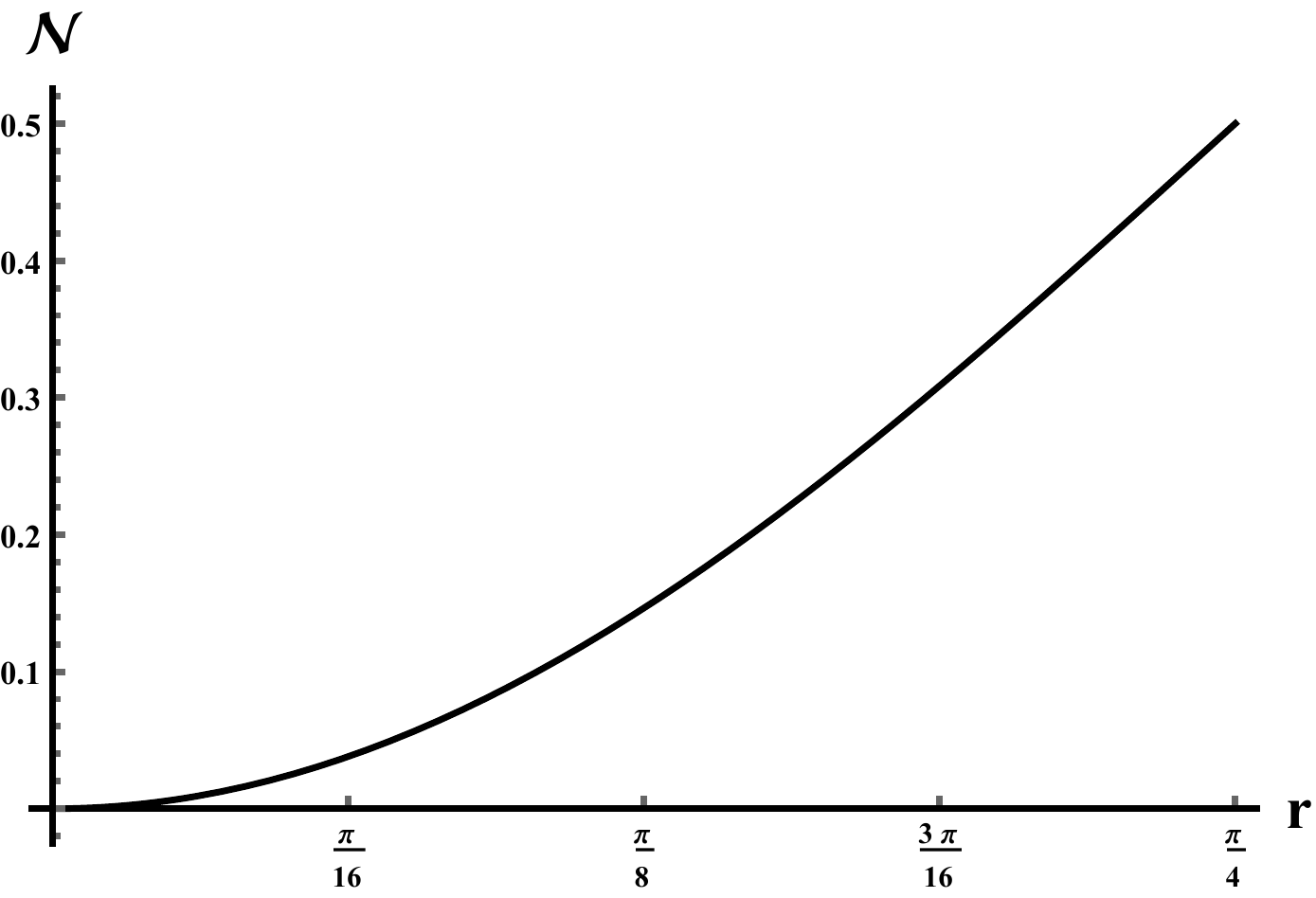}
		\label{FGWS9fig:negativity12}
	\end{subfigure}
	~
	\begin{subfigure}[t]{0.5\textwidth}
		\centering
		\caption{}
		\includegraphics[width=3in,height=3in]{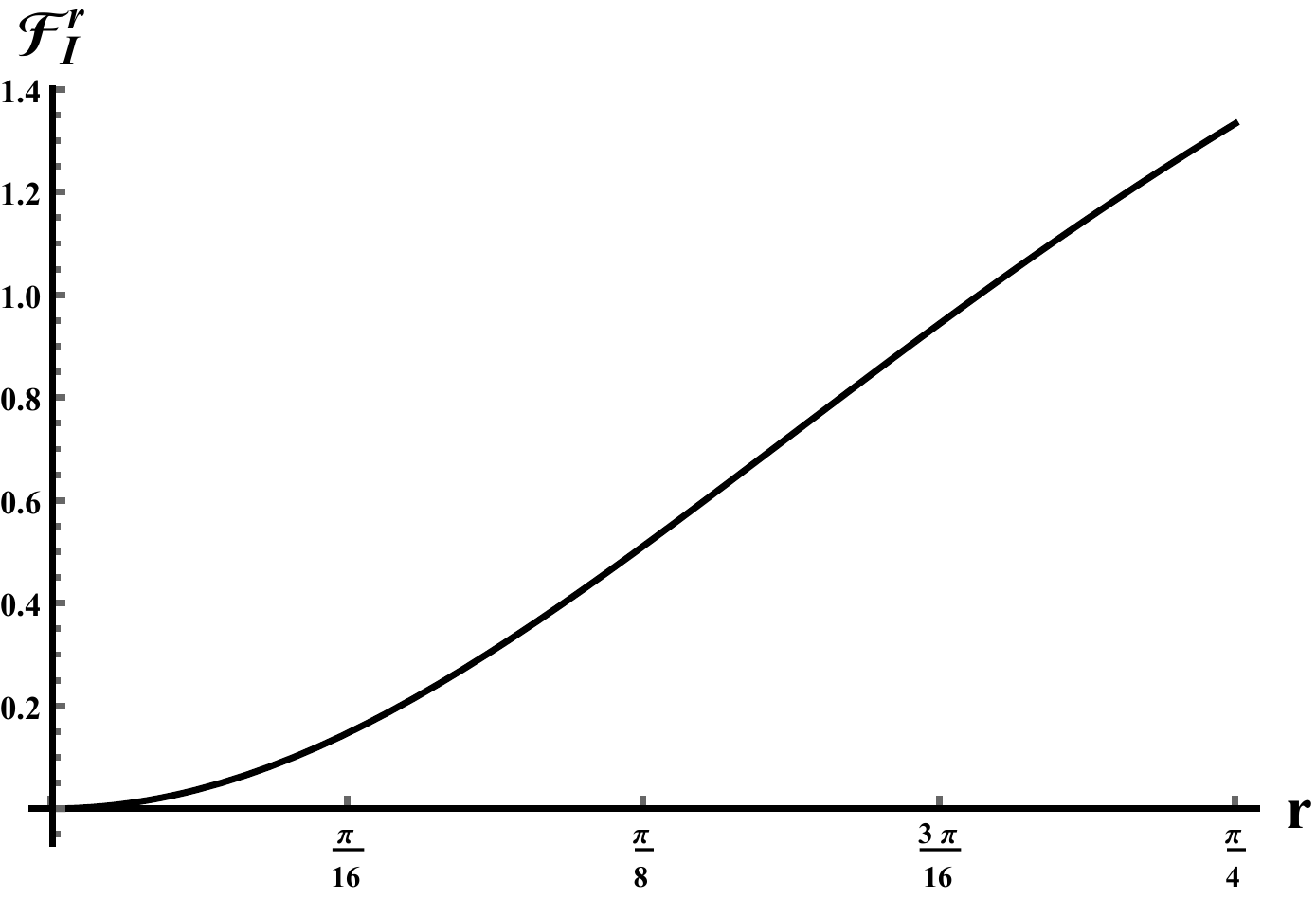}
		\label{FGWS9fig:QFaftermaskingr12}
	\end{subfigure}
	\caption{For the  partition \eqref{FGWS9rho12}, we evaluate (\subref{FGWS9fig:negativity12}) the negativity for the reduced density operator $\rho_{12}(\rho_{34})$  (\subref{FGWS9fig:QFaftermaskingr12}) The quantum Fisher information $\mathcal{F}_{I}^r$}
	\label{FGWS9fig:N2}
\end{figure}
}
Finally the local and the non-local information that encoded on the partition  \eqref{FGWS9rho12}  are shown in \figurename{(\ref{FGWS9fig12:figureloc})}.  Again since the accelerated partition  \eqref{FGWS9rho12}  depends only on the acceleration $r$, and its  amount of the quantum correlation  decreases as $r$ increases, then the amount of non-local information deceases as $r$ increases. Therefore, one may conclude that, at  any values of the initial states settings, one can find a partition depends only on the accelerated parameters, where this partition satisfies the masking conditions. However, the behavior of the non-local information encoded in the state   \eqref{FGWS9rho12}  is displayed in  \figurename{(\ref{FGWS9fig12:nlocbefmask})}. However, the behavior of the $I_{nlocal}$ shows that it decays slightly as $r$ increases. This is due to the robustness of the partition  \eqref{FGWS9rho12} against the acceleration, where as it is displayed from \figurename{(\ref{FGWS9fig:QFaftermaskingr12})}, the entanglement increases as $r$ increases.

 The most important result that is  depicted in \figurename{(\ref{FGWS9fig12:nlocbefmask})} for the local information. It seems that, the local information  $I_{loc}$  is independent of the acceleration parameter, $r$. By comparing  \figurename{(\ref{FGWS9fig:locbemask})}  and \figurename{(\ref{FGWS9fig12:locbemask})}, one can notice that, before applying the  masking process the local information decreases as $r$ increases, while after the masking process  it is independent of the acceleration parameter. However, these results can be obtained analytically  from the final state of the first qubit $\rho_1=\tr_2\{\rho_{12}\}=\frac{1}{2}I_2$. This means that, the local information is hidden within the non-local quantum correlation.

{\normalsize
\begin{figure}[t!]
	\begin{subfigure}[t]{0.5\textwidth}
		\centering
		\caption{}
		\includegraphics[width=3in,height=3in]{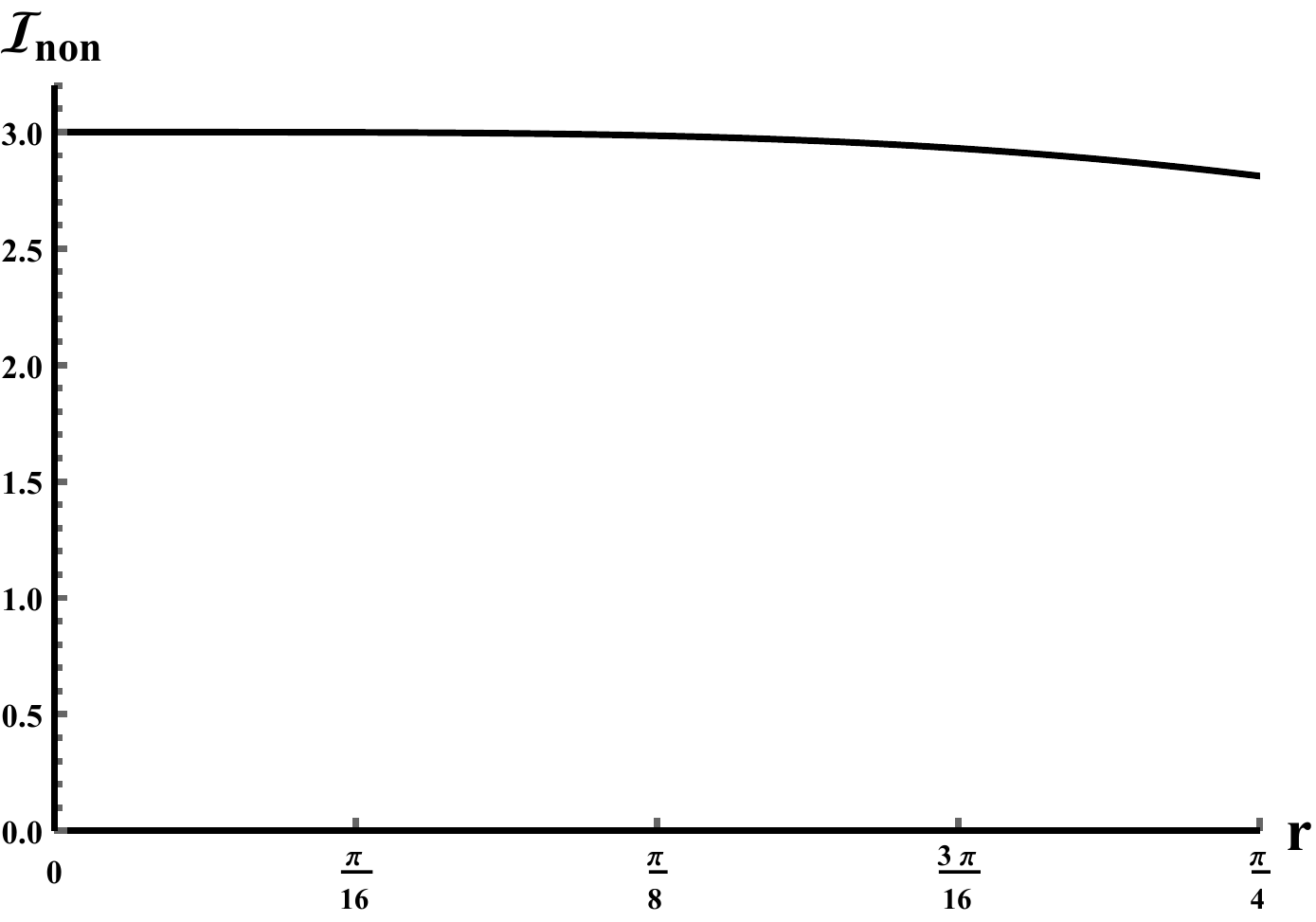}
		\label{FGWS9fig12:nlocbefmask}
	\end{subfigure}
	~
	\begin{subfigure}[t]{0.5\textwidth}
		\centering
		\caption{}
		\includegraphics[width=3in,height=3in]{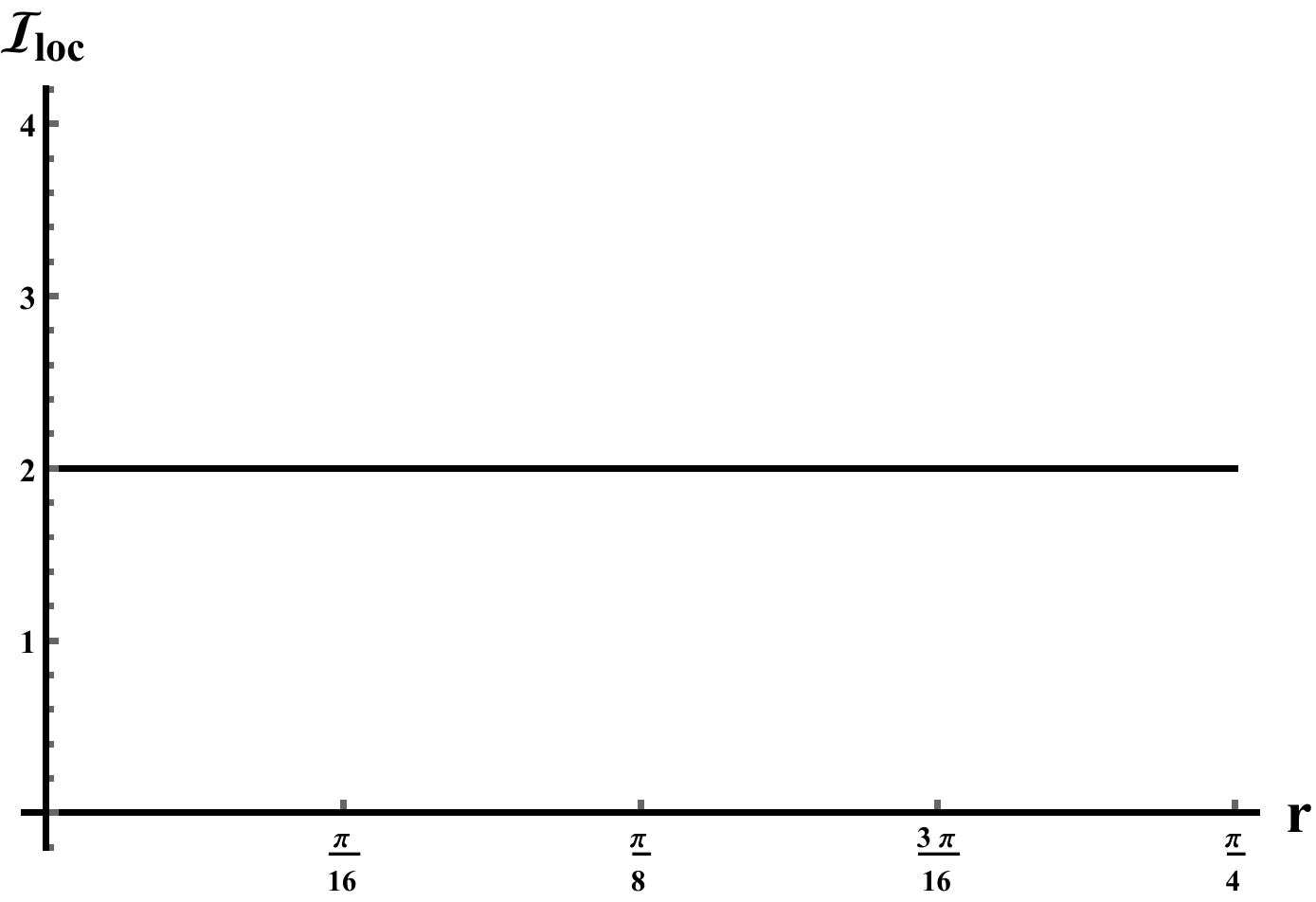}
		\label{FGWS9fig12:locbemask}
	\end{subfigure}
	\caption{ The local and non-local information for \eqref{FGWS9rho12}}
	\label{FGWS9fig12:figureloc}
\end{figure}
}


On the other hand, by using the PPT \cite{Peres1996} criterion, one can  find that the eigenvalues of  $\rho_{13}^{T_2}$ are non-negative, and consequently the state  \eqref{FGWS9rho13} is separable, namely contains only classical correlation on the range of its settings, where $|\eta|\leq 1$, where $\eta=x,y,z$. Meanwhile its subsystems satisfy the  conditions of masking.
Similarly, we  examine the behavior of the  classical Fisher information (CFI) and local/non-local information that are  masked in  the second partition \eqref{FGWS9rho13}.
In this context, the classical Fisher information $\mathcal{F}_C^{r}$ and $\mathcal{F}_C^x$ are obtained analytically as,
\begin{equation}
\mathcal{F}_C^r=\frac{x^2 \sin ^2(2 r)}{x^2 \cos ^4 r-1}, \quad \quad \mathcal{F}_C^x=\frac{\cos ^4(r)}{1-x^2 \cos ^4 r}.
\label{CFIrho13}
\end{equation}
From \eqref{CFIrho13}, the singularity of the two functions are depicted at $x=\pm 1$ and $r=0$. Therefore mathematically the choice of $x=\pm1$ is not to be considered.

\begin{figure}[t!]
	\begin{subfigure}[t]{0.5\textwidth}
		\centering
		\caption{}
		\includegraphics[width=3in,height=3in]{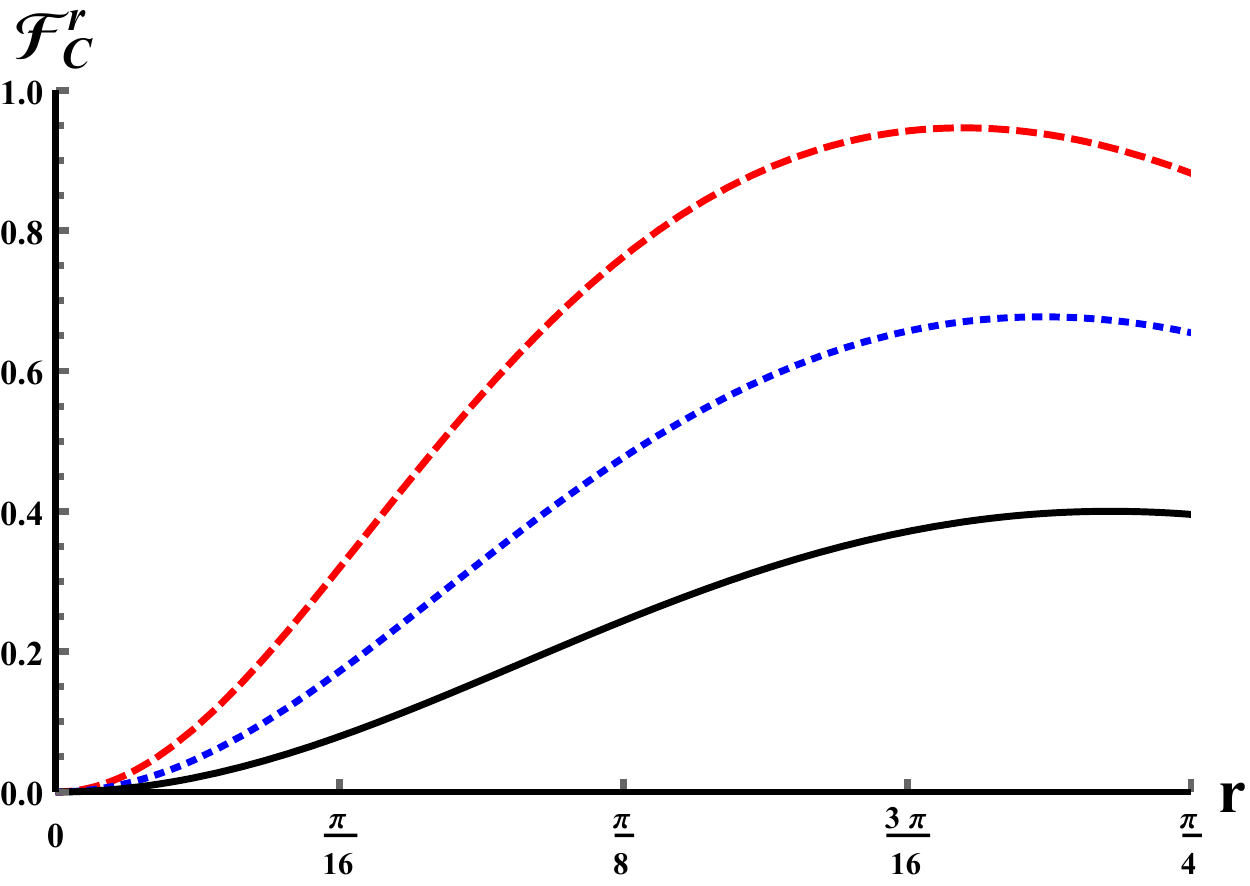}\label{FGWS9fig:QFaftermaskingr}
	\end{subfigure}
	~
	\begin{subfigure}[t]{0.5\textwidth}
		\centering
		\caption{}
		\includegraphics[width=3in,height=3in]{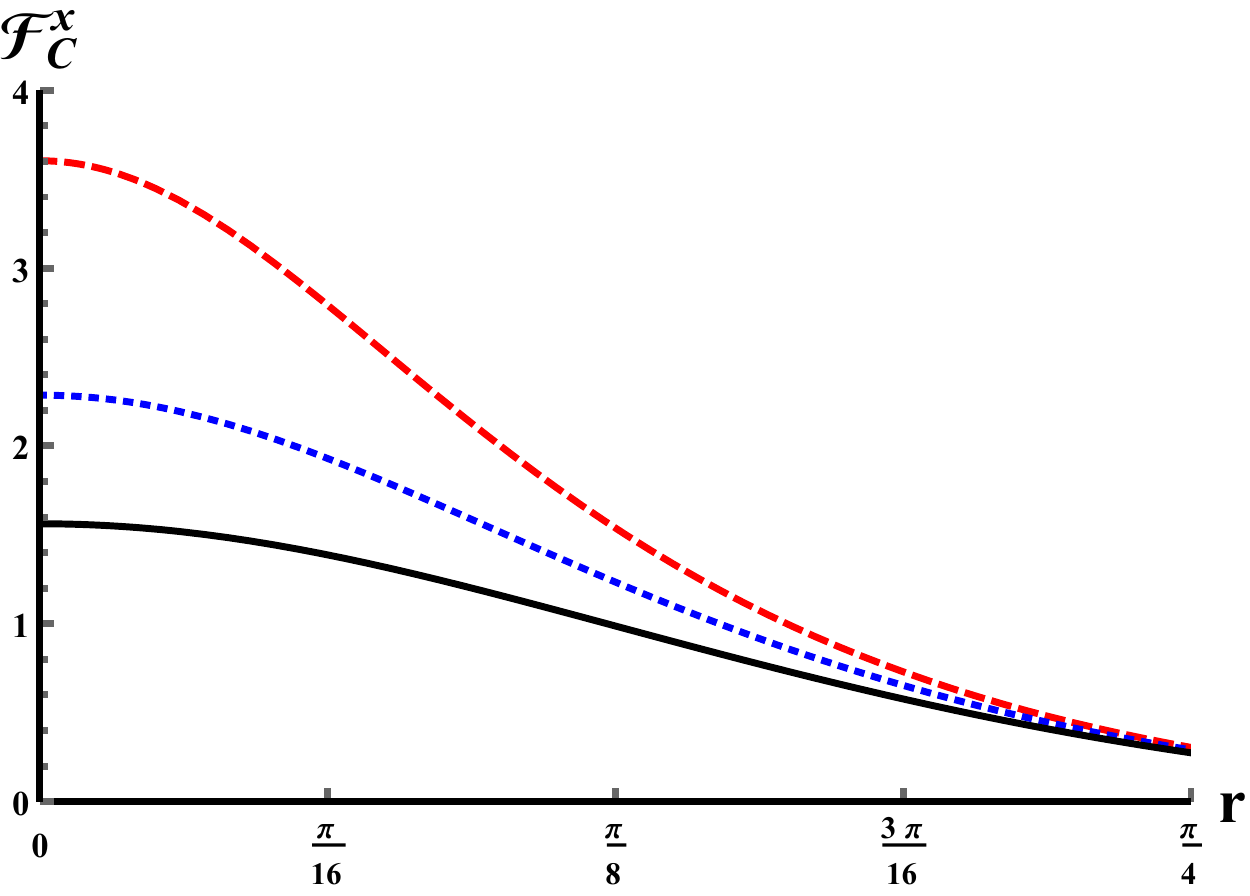}
		\label{FGWS9fig:QFaftermaskingx}
	\end{subfigure}
	\caption{ Estimating  (\subref{FGWS9fig:QFaftermaskingr}) the acceleration parameter $r$, and (\subref{FGWS9fig:QFaftermaskingx}) the $x$ parameter by using  \eqref{FGWS9rho13}. It is assumed that the system is initially prepared in a Werner states  with $x = y = z = -0.85$ for the dashed curve, $x = y = z = -0.75$ for the dotted curve and the $X$- state with $x = - 0.6$, $y = - 0.2$ and $z = - 0.5$ for the solid curve.}
	\label{FGWS9fig:figure13maskr1}
\end{figure}

The accelerated  masked state \eqref{FGWS9rho13} is a function of the parameters $r$ and $x$.  Therefore, we can estimate these parameters by means of the classical Fisher information(CFI). The behaviors of CFI with respect to the parameter $r$ is shown in \figurename{(\ref{FGWS9fig:QFaftermaskingr})}. It is clear that, $\mathcal{F}_C^r$ increases as the acceleration increases. However, as one increases  initial entanglement of the acceleration system, the possibility of estimating $r$ classically increases.    However,  at other values of the acceleration, the classical  Fisher information $\mathcal{F}_C^r$ reaches its maximum values at different acceleration.

The classical Fisher information with respect to the parameter $x$ is explored in  \figurename{(\ref{FGWS9fig:QFaftermaskingx})}. Since, the initial state settings demonstrates  the degree of entanglement and the masked state \eqref{FGWS9rho13} is a product state and consequently its separability increasers as $r$ increases. Therefore the estimation degree of the parameter $x$ decreases as $r$ increases. These results can be seen from the analytical form \eqref{CFIrho13}.

It is worth  to mention that, one of the most important advantages of the masking process in this case is: it is possible to mask any state of accelerated two-qubit system, where the masked state either depends on only the acceleration parameter as $\rho^{mask_{ab}}_{12(34)}$ or  the acceleration parameter $r$ and only one parameter $x$, where $|x|\leq 1$ and the other parameters are arbitrary. Also, as it is displayed from the non-local product state \eqref{FGWS9rho13}, the users can not estimate any parameter without their contribution. This technique represents a kind of protecting the information.

\begin{figure}[t!]
	\begin{subfigure}[t]{0.5\textwidth}
		\centering
		\caption{}
		\includegraphics[width=3in,height=3in]{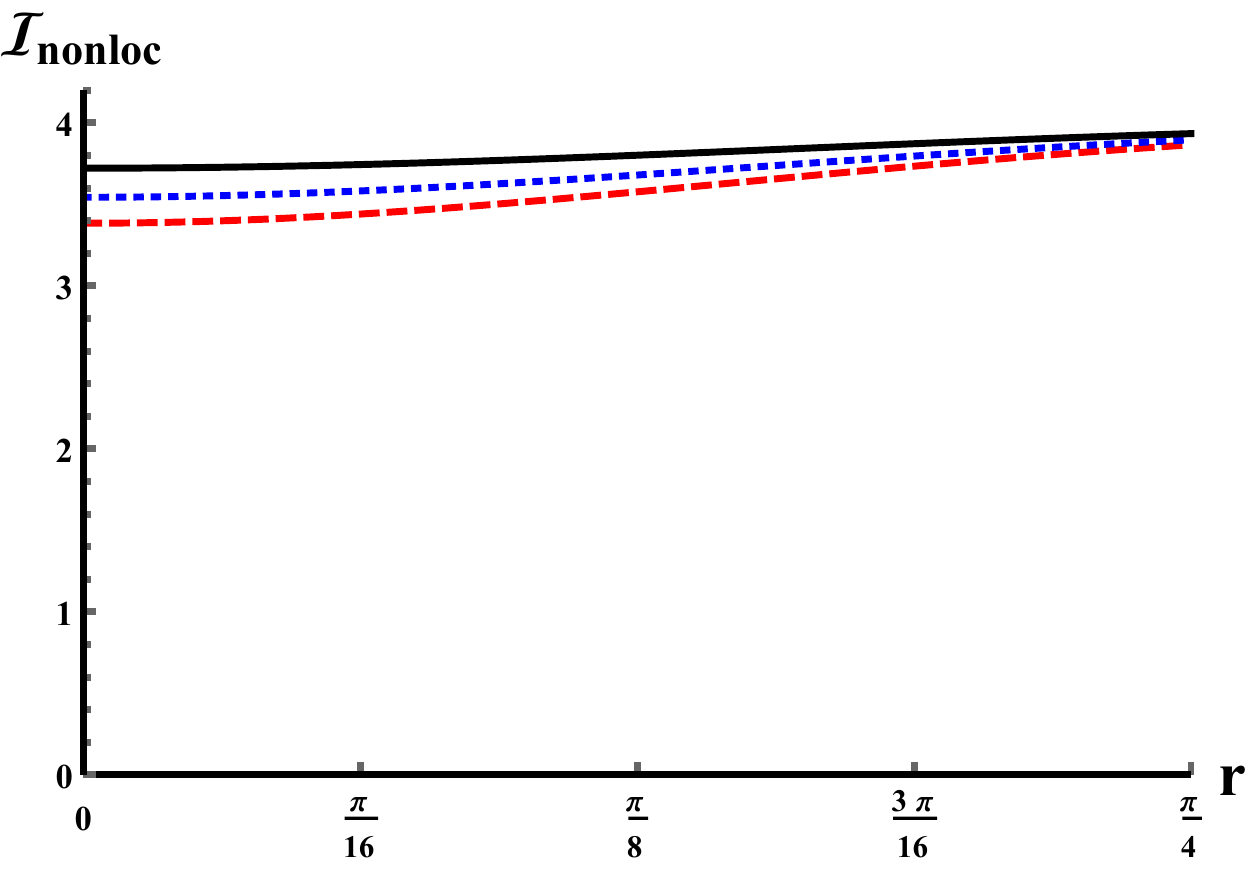}
		\label{FGWS9fig13:nlocbefmask}
	\end{subfigure}
	~
	\begin{subfigure}[t]{0.5\textwidth}
		\centering
		\caption{}
		\includegraphics[width=3in,height=3in]{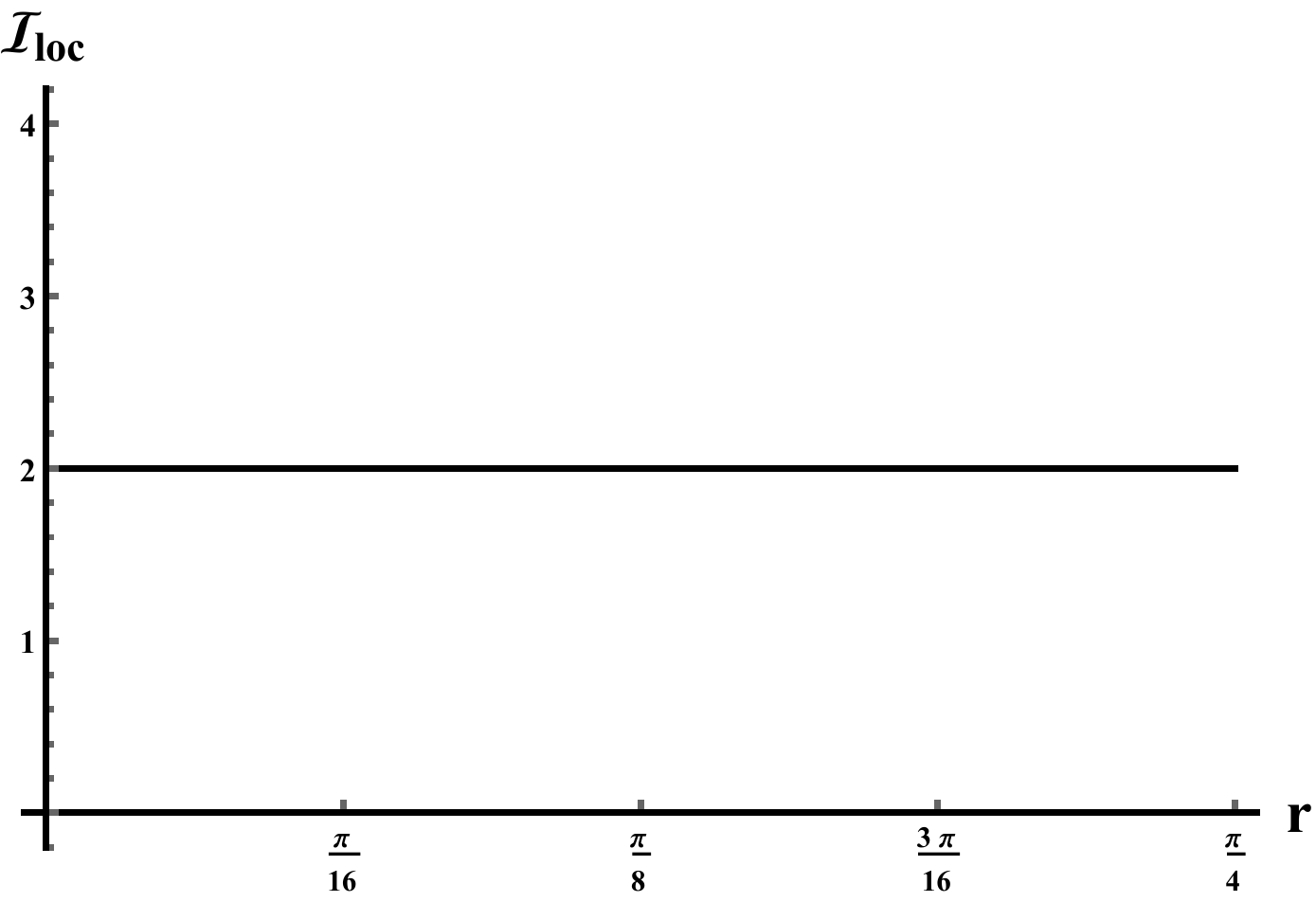}
		\label{FGWS9fig13:locbemask}
	\end{subfigure}
	\caption{ The local and non-local information for \eqref{FGWS9rho13}}
	\label{FGWS9fig13:figureloc}
\end{figure}

To look at the effect of the masking process on the behavior of the  local and non-local information of the masked state \eqref{FGWS9rho13}, we  consider  the initial accelerated state to be prepared in different state settings.  As it is displayed from \figurename{(\ref{FGWS9fig13:nlocbefmask})}, the $\mathcal{I}_{nloc}$  increases as the acceleration parameter $r$ increases. Moreover, the largest values are predicted for an accelerated system initially prepared in the $X$-state. The non-local information slightly increases as $r$ increases, where  the increasing rate is displayed clearly for those initially  encoded the large entangled state.  The behavior of the local information is displayed in \figurename{(\ref{FGWS9fig13:locbemask})}, is independent of the acceleration $r$. On the other hand, by comparing \figurename{(\ref{FGWS9fig13:nlocbefmask})} and \figurename{(\ref{FGWS9fig:nlocbefmask})}, we can see that the amount of the non-local information that displayed in \figurename{(\ref{FGWS9fig13:nlocbefmask})} is much larger than that shown in \figurename{(\ref{FGWS9fig:nlocbefmask})}. This means that the local information is coded now in the non-local classical correlation.\\
These results display that, the local encoded information can be  quantified from  both qubits, the masker qubit which is in Alice hand and one of Bob's qubits which may be the masker qubit or the initial accelerated qubit.  Therefore, one can not obtain any information from Alice's qubit alone.

\subsection{  Only Alice's qubit is masked}
\label{Both-accelereted-Alice-masked}
In this subsection, it is assumed that both qubits accelerated and only Alice's qubit is masked. Therefore, we get

\begin{align}
\rho^{\text{acc}}_{123}&=\frac{1}{4}\left\lbrace\right.
\cos ^2r \left( \ket{000} \bra{000} + \ket{110} \bra{110}\right)
\\&\quad\quad+
x \cos ^2r \left( \ket{001}\bra{000} + \ket{000}\bra{001} - \ket{111}\bra{110} - \ket{110}\bra{111} \right)
\\&\quad\quad+
\cos ^2r \left( z \cos ^2r-\sin ^2r \right) \left(  \ket{110}\bra{000} + \ket{000}\bra{110} \right)
\\&\quad\quad+
y \cos ^2r \left( \ket{111}\bra{000} - \ket{110}\bra{001} - \ket{001}\bra{110} + \ket{000}\bra{111} \right)
\\&\quad\quad
+\left( 2- \cos ^2 r\right) \left( \ket{001}\bra{001} +\ket{111}\bra{111} \right)
\\&\quad\quad
+(3\cos ^2 r -(z+1) \cos ^4 r-2)  \left( \ket{111}\bra{001} +\ket{001}\bra{111} \right)
\left.\right\} .
\end{align}

From this density operator, we have three partitions $\rho^{mask_a}_{12}=tr_3\{\rho^{mask_a}_{123}\}$,    $\rho^{mask_a}_{13}=tr_2\{\rho^{mask_a}_{123}\}$, and   $\rho^{mask_a}_{23}=tr_1\{\rho^{mask}_{123}\}$. In the computational basis the partition $\rho^{mask_a}_{12}$ can be written explicitly as,

\begin{align}
\rho_{12}^{mask_a}&=\frac{1}{2}\left( \ket{00} \bra{00} + \ket{11} \bra{11} \right) -\frac{1}{2}\sin^2 r \left(  \ket{11}\bra{00} + \ket{00}\bra{11} \right).
\label{FGWS7rho12}
\end{align}

It is clear that, the state (\ref{FGWS7rho12}),  satisfies the masking conditions, where $\rho_a^{mask_a}=tr_2\{\rho_{12}^{mask_a}\}=\frac{1}{2}I_{2}$ and   $\rho_2^{mask}=tr_1\{\rho_{12}^{mask}\}=\frac{1}{2}I_2 $. However,the state  (\ref{FGWS7rho12}) is the same as \eqref{FGWS9rho12}. Therefore it predicts the same behavior of the QFI, local and non-local information.
The second and third partitions are equal, explicitly they may be written as,

\begin{align}
\rho^{mask_a}_{13}&=\rho^{mask_a}_{23} \\
&=\frac{1}{4}\left\lbrace\right.
\cos ^2 r\left( \ket{00}\bra{00}+\ket{10}\bra{10}\right)+ (2 -  \cos ^2 r) \left(  \ket{01}\bra{01} + \ket{11}\bra{11} \right)
\\& \quad + x \cos ^2 r\left( \ket{00}\bra{01} + \ket{01}\bra{00} - \ket{10}\bra{11} - \ket{11}\bra{10} \right)
\left.\right\} .
\label{FGWS7rho13}
\end{align}

The reduced density operators of the subsystems  are defined such that, $\rho_1^{mask_a}=\rho_3^{mask_a}=\frac{1}{2}I_{2\times 2}$.
This means that  the state \eqref{FGWS7rho13} satisfies the masking condition.
On the other hand, the  masked state  \eqref{FGWS7rho13} is separable, where the eigenvalues of $ (\rho^{mask_a}_{13})^{T_2}$ are non-negative. Therefore the local information is masked on the classical correlation.

It is clear that the masked  state \eqref{FGWS7rho13} is a function of the parameters $r$ and $x$.  Therefore, we can estimate these parameters by means of CFI, which requires evaluating  the eigenvalues  as given by,
\begin{align}
&\lambda_{1}=\lambda_{2}=\frac{1}{4} \left(1-\kappa (x,y)\right), \quad\quad
\lambda_{3}=\lambda_{4}=\frac{1}{4} \left(1+\kappa (x,y)\right),
\end{align}
where, $\kappa (x,y)=\sqrt{\left(x^2+1\right) \cos ^4(r)-2 \cos ^2(r)+1}$. The analytical expressions of the  classical Fisher information of $\mathcal{F}^{r}_C$ and $\mathcal{F}^x_C$ are given by,
\begin{equation}
\mathcal{F}_C^r=\frac{\sin ^2(2 r) \left(\left(x^2+1\right) \cos ^2  r-1\right)^2}{ \kappa^2 (1-\kappa^2)}, \quad \quad \mathcal{F}_C^x=\frac{x^2 \cos ^8 r}{\kappa^2 (1-\kappa^2)}.
\label{CFI-A}
\end{equation}

\begin{figure}[t!]
	\begin{subfigure}[t]{0.5\textwidth}
		\centering
		\caption{}
		\includegraphics[width=3in,height=3in]{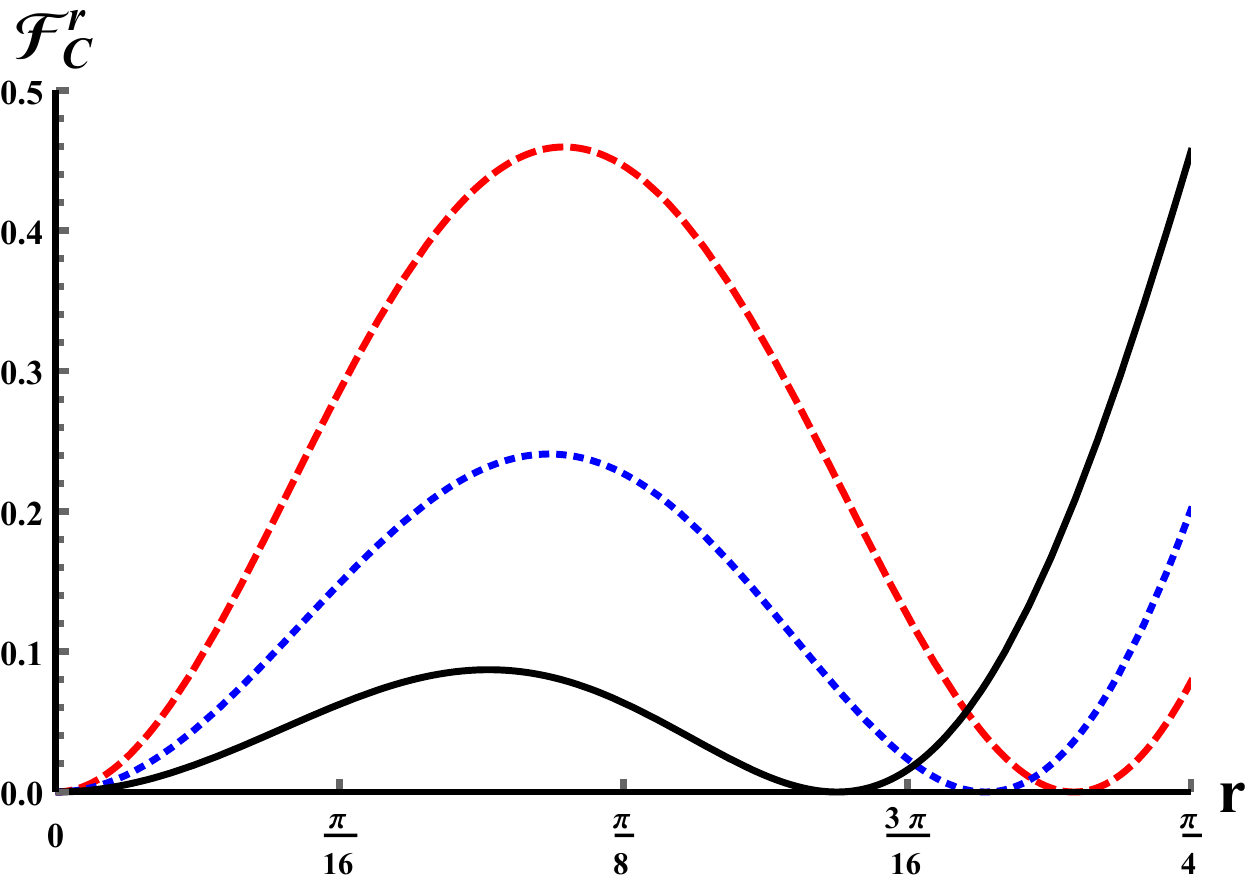}
	
		\label{FGWS7fig:QFaftermaskingr}
	\end{subfigure}
	~
	\begin{subfigure}[t]{0.5\textwidth}
		\centering
		\caption{}
		\includegraphics[width=3in,height=3in]{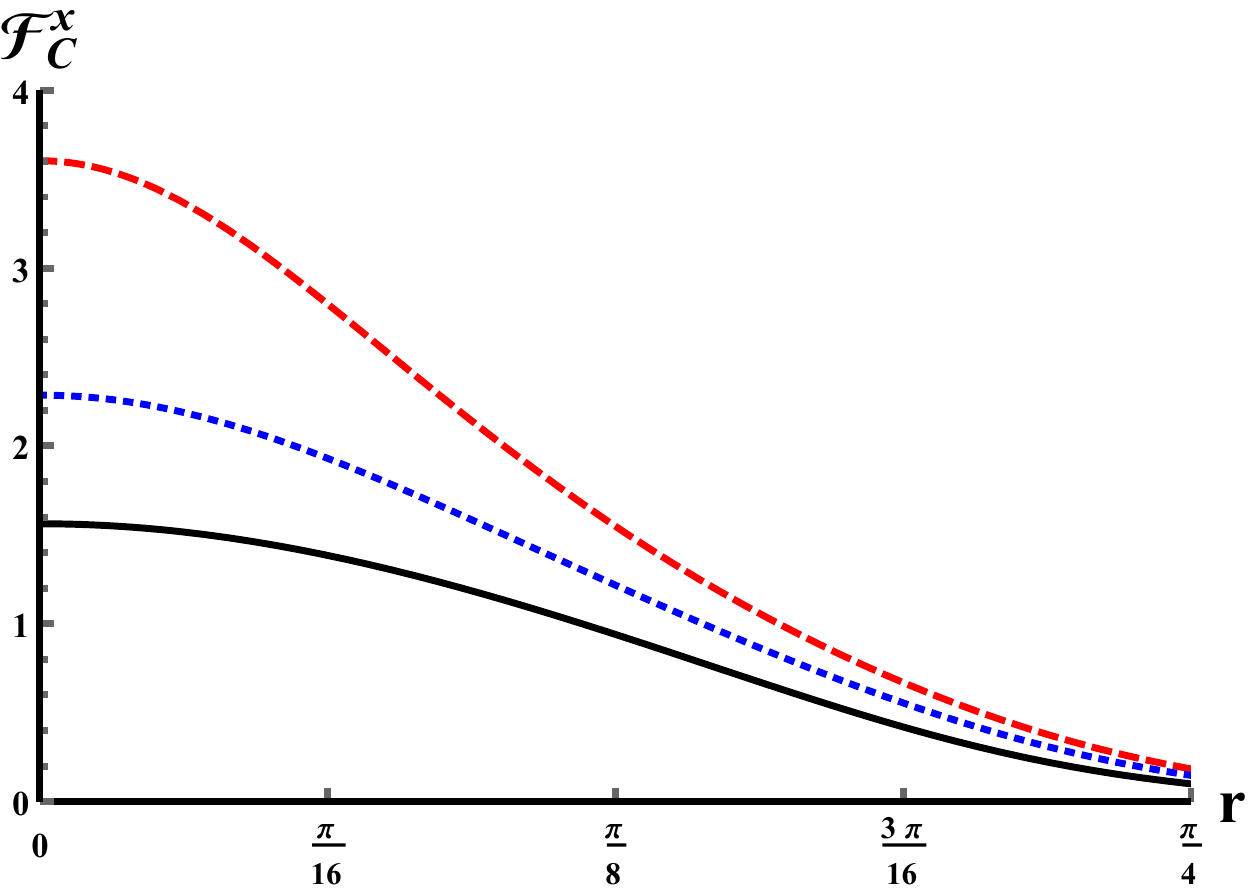}\label{FGWS7fig:QFaftermaskingx}
	\end{subfigure}
	\caption{ The same as  \figurename{(\ref{FGWS9fig:figure13maskr1})} but for the masked state  \eqref{FGWS7rho13}.}
	\label{FGWS7fig:figure13maskr1}
\end{figure}

As it is exhibited from the  mathematical form \eqref{CFI-A}, $\mathcal{F}_{C}^{r}$ has the minimum value of zero at $r=0$ and $r=\arccos(\pm\frac{1}{\sqrt{x^2+1}})$,  while  $\mathcal{F}_{x}^{r}$ has no singular point on the domain of $x\in(-1,1)$.
 The behavior of CFI with respect to the parameter $r$ and $x$ is shown in \figurename{(\ref{FGWS7fig:figure13maskr1})}, where we plot $\mathcal{F}_{C}^{r}$ for different initial state settings.
 It is clear that, $\mathcal{F}_{C}^{r}$ increases gradually as $r$ increases to reach its maximum  bound. After that decreases gradually to vanish completely at $r=\arccos(\pm\frac{1}{\sqrt{x^2+1}})$. However, at further values of $r$, it  increases again to reach its maximum value at $r=\pi/4$. The behavior of the classical Fisher information $\mathcal{F}^{x}_C$ is shown in  \figurename{(\ref{FGWS7fig:QFaftermaskingx})}, where it decreases gradually to reach their minimum values at $r=\pi/4$.  The minimum bounds of $\mathcal{F}^{x}_C$ depend on the initial entanglement of the accelerated state.


\begin{figure}[t!]
	\begin{subfigure}[t]{0.5\textwidth}
		\centering
		\caption{}
		\includegraphics[width=3in,height=3in]{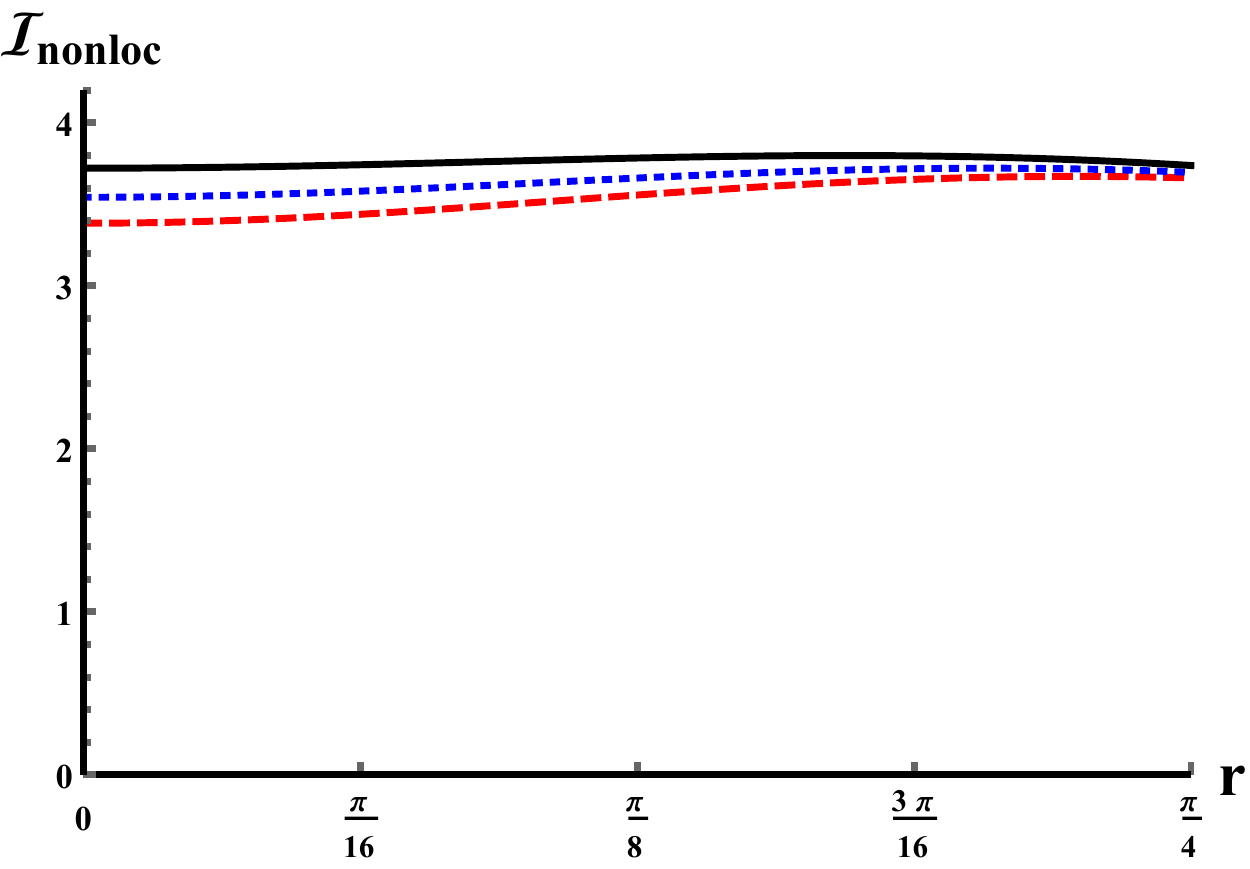}
		\label{FGWS7fig13:nlocbefmask}
	\end{subfigure}
	~
	\begin{subfigure}[t]{0.5\textwidth}
		\centering
		\caption{}
		\includegraphics[width=3in,height=3in]{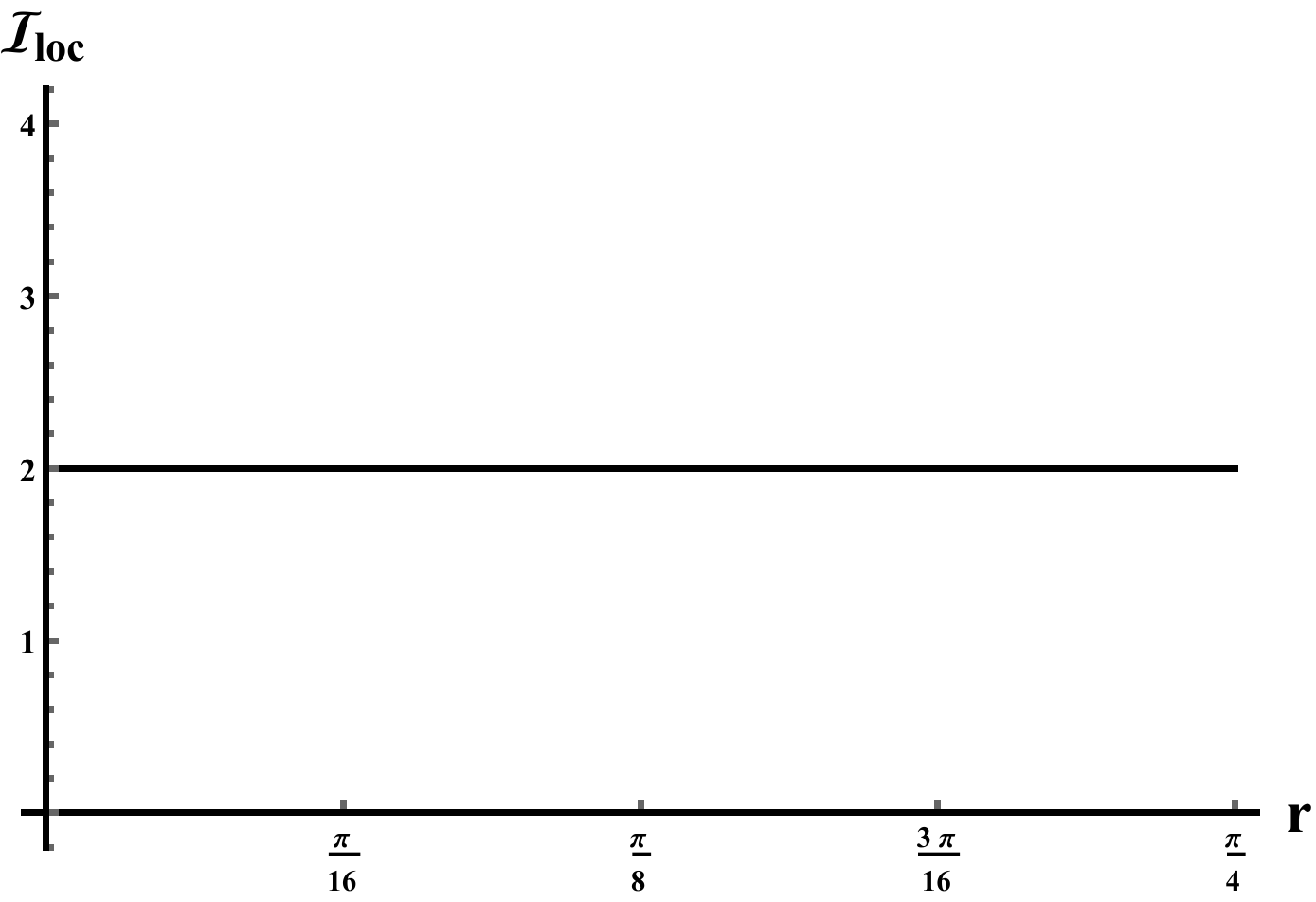}
		\label{FGWS7fig13:locbemask}
	\end{subfigure}
	\caption{The local and non-local information for \eqref{FGWS7rho13}.}
	\label{FGWS7fig13:figureloc}
\end{figure}

The local and non-local information of the state \eqref{FGWS7rho13}  are shown in \figurename{(\ref{FGWS7fig13:figureloc})}. The  behavior of $\mathcal{I}_{nloc}$ is similar to that displayed in \figurename{(\ref{FGWS9fig13:nlocbefmask})}, but the non-local information is slightly smaller.  Moreover, the maximum bound is much larger than those displayed before masking process, because the masked local information are also encoded on the classical correlation.

\section{Alice is accelerated}
\label{Alice-accelerated}
In this section, we assume that only Alice qubit is accelerated. Using the acceleration process, the final  accelerated state is given by,
\begin{align}
\rho_{\text{ac}}&= \mathcal{B}_{11} \ket{00}\bra{00}+\mathcal{B}_{14} \ket{00}\bra{11}+\mathcal{B}_{22} \ket{01}\bra{01} + \mathcal{B}_{23} \ket{01}\bra{10} \\& \quad+\mathcal{B}_{32}\ket{10}\bra{01} +\mathcal{B}_{33} \ket{10}\bra{10}
 +\mathcal{B}_{41}\ket{11}\bra{00}+\mathcal{B}_{44} \ket{11}\bra{11},\\
\label{FGWS1rhoaccelerated}
\end{align}
where
\begin{align}
&\mathcal{B}_{11} = c^2 \mathcal{A}_{11} ,  \quad
\mathcal{B}_{14} = c \mathcal{A}_{14} ,  \quad
\mathcal{B}_{22} = c^2 \mathcal{A}_{22},  \quad
\mathcal{B}_{23} = c \mathcal{A}_{23}  ,  \quad
\mathcal{B}_{32} = c \mathcal{A}_{32}  ,  \quad\\&
\mathcal{B}_{33} = s^2 \mathcal{A}_{11} + \mathcal{A}_{33} ,  \quad
\mathcal{B}_{41} = c \mathcal{A}_{41}  ,  \quad
\mathcal{B}_{44} = s^2 \mathcal{A}_{22} +\mathcal{A}_{44}.
\end{align}
The behavior of the amount of the non-classical correlations (entanglement) contained in the state \eqref{FGWS1rhoaccelerated},  is   examined through the negativity in  \figurename{(\ref{FGWS1fig:negativityacc})}, where different initial state settings are considered.
{\normalsize
\begin{figure}[H]
	\centering
	{\includegraphics[width=3in,height=3in]{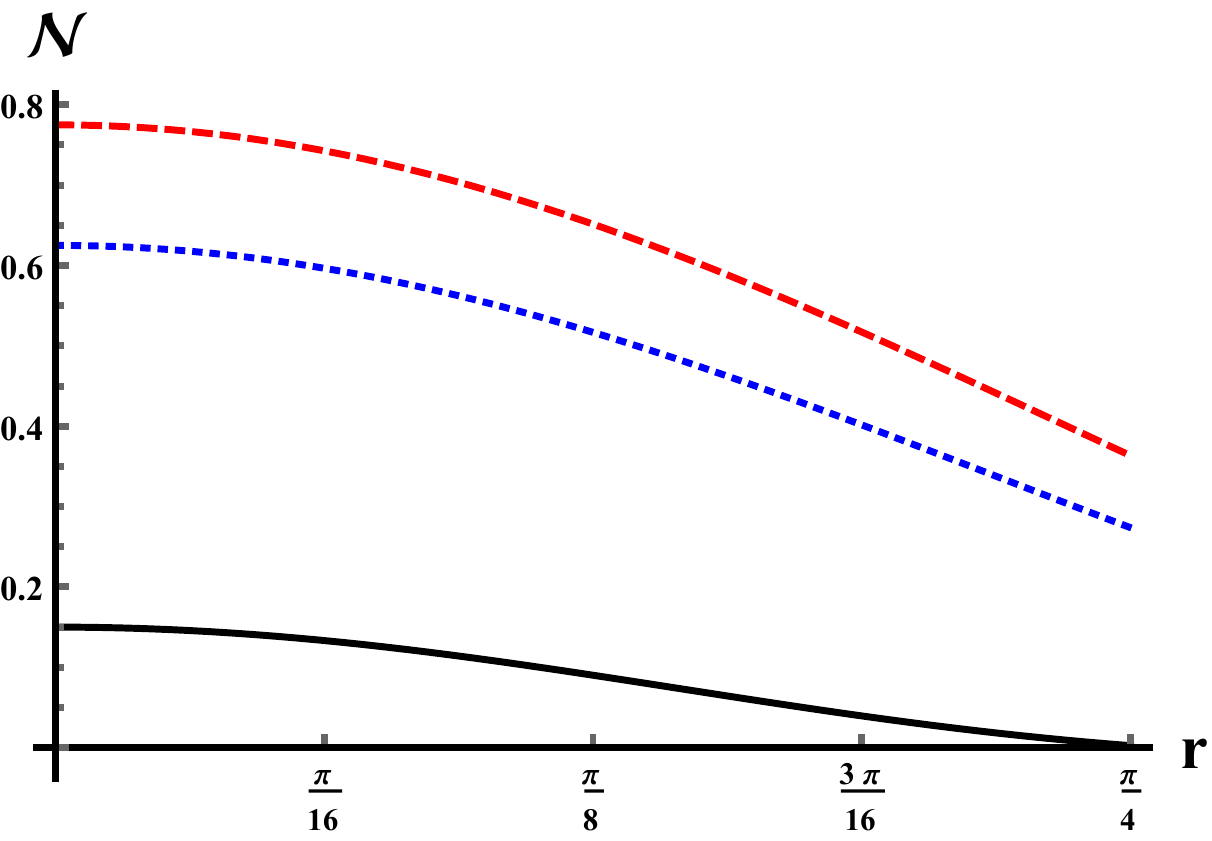}}
	\caption{The  negativity for the state \eqref{FGWS1rhoaccelerated}, where it is assumed that only Alice's qubit is accelerated. The system is initially prepared in a Werner state with  $x = y=z=-0.85$, and $x=y=z=-0.75$ for the dashed  and the dot curves, respectively. While the solid curve represents the behavior of the negativity for the $X$-state with $x = - 0.6$, $y = - 0.2$ and $z = - 0.5$.}
	\label{FGWS1fig:negativityacc}
\end{figure}
}
The behavior of the negativity, as a measure of the entanglement, is  similar to that shown in \figurename{(\ref{FGWS9fig:negativityacc})}, namely it decreases as the acceleration increase. The decreasing rate depends on the initial entanglement of the accelerated system. Moreover, the decreasing rate is much smaller than that displayed when both qubits are accelerated, as shown from \figurename{(\ref{FGWS9fig:negativityacc})}. The negativity decreases faster and vanishes at small values of $r$ for the suggested $X$-state settings.
Moreover, the accelerated state  \eqref{FGWS1rhoaccelerated} is a function of  the acceleration parameter and the one that describes the initial accelerated state.  Therefore for the sake of matching we estimate the parameters $r$ and $x$. \figurename{(\ref{FGWSfig:figure1r1}a)} displays the behavior of classical Fisher information (CFI) with respected to the acceleration $r$, $\mathcal{F}^r_C$. However, as $r$ increase, $\mathcal{F}^r_C$ increases gradually to reach its maximum value. At further values of $r$, $\mathcal{F}^r_C$, decreases gradually to reach its minimum values at $r=\pi/4$.
The parameter $x$ is estimated by using the classical Fisher information $\mathcal{F}^x_{C}$ for different initial state settings see (\figurename{(\ref{FGWSfig:figure1r1}b)}). In this case, $\mathcal{F}^x_{C}$ decreases as $r$ increases. The largest value of the classical Fisher information $\mathcal{F}^x_{C}$ is depicted for the less initial entangled  accelerated state at small values of $r$.

\begin{figure}[t!]
	\begin{subfigure}[t]{0.5\textwidth}
		\centering
		\caption{}
		\includegraphics[width=3in,height=3in]{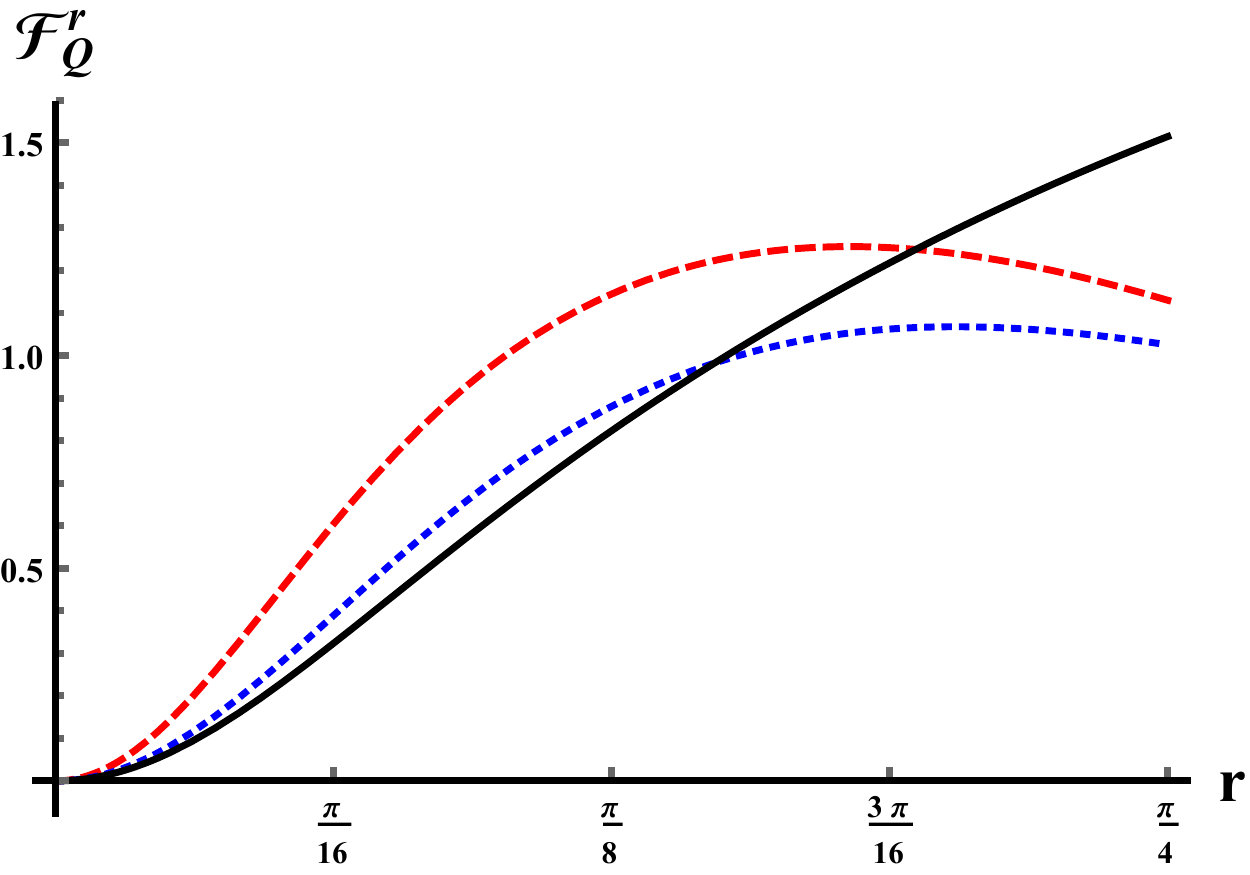}
		\label{FGWS1fig:QFbeforemaskingr}
	\end{subfigure}
	~
	\begin{subfigure}[t]{0.5\textwidth}
		\centering
		\caption{}
		\includegraphics[width=3in,height=3in]{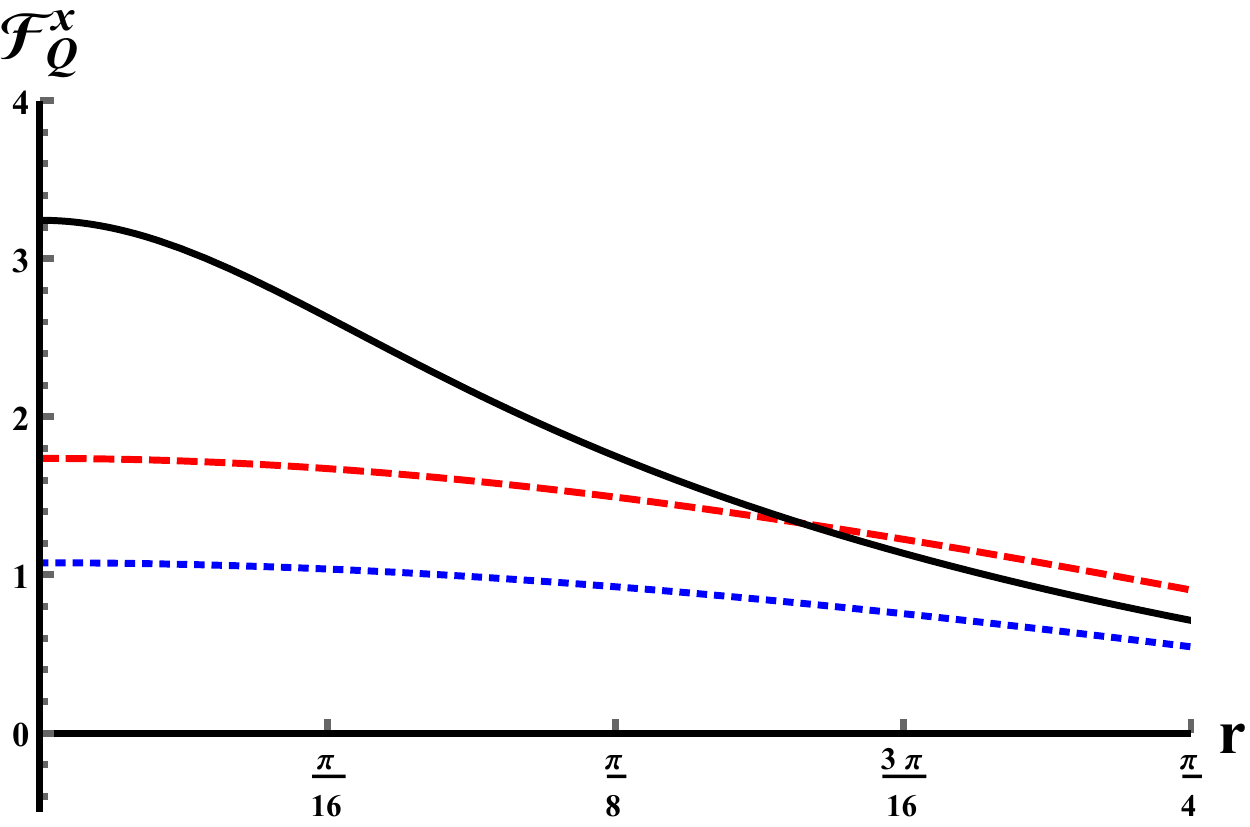}
		\label{FGWS1fig:QFbeforemaskingx}
	\end{subfigure}
	\caption{ Estimating  (\subref{FGWS1fig:QFbeforemaskingr}) the acceleration parameter $r$, and (\subref{FGWS1fig:QFbeforemaskingx}) the $x$ parameter by using  \eqref{FGWS1rhoaccelerated}. It is assumed that the system is initially prepared in a Werner states  with $x = y = z = -0.85$ for the dashed curve, $x = y = z = -0.75$ for the dotted curve and the $X$- state with $x = - 0.6$, $y = - 0.2$ and $z = - 0.5$ for the solid curve.}
	\label{FGWSfig:figure1r1}
\end{figure}

\begin{figure}[t!]
	\begin{subfigure}[t]{0.5\textwidth}
		\centering
		\caption{}
		\includegraphics[width=3in,height=3in]{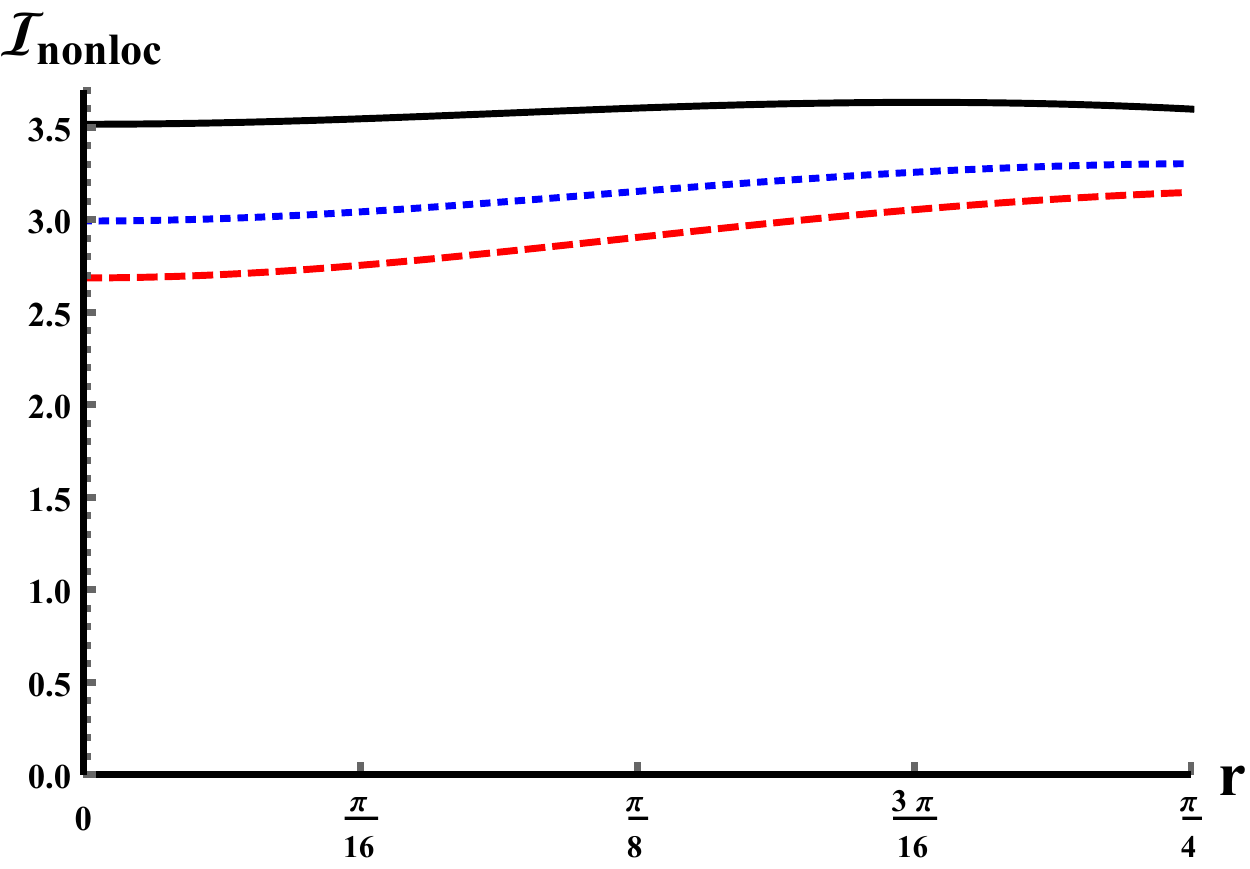}
		\label{FGWS1fig:nlocbefmask}
	\end{subfigure}
	~
	\begin{subfigure}[t]{0.5\textwidth}
		\centering
		\caption{}
		\includegraphics[width=3in,height=3in]{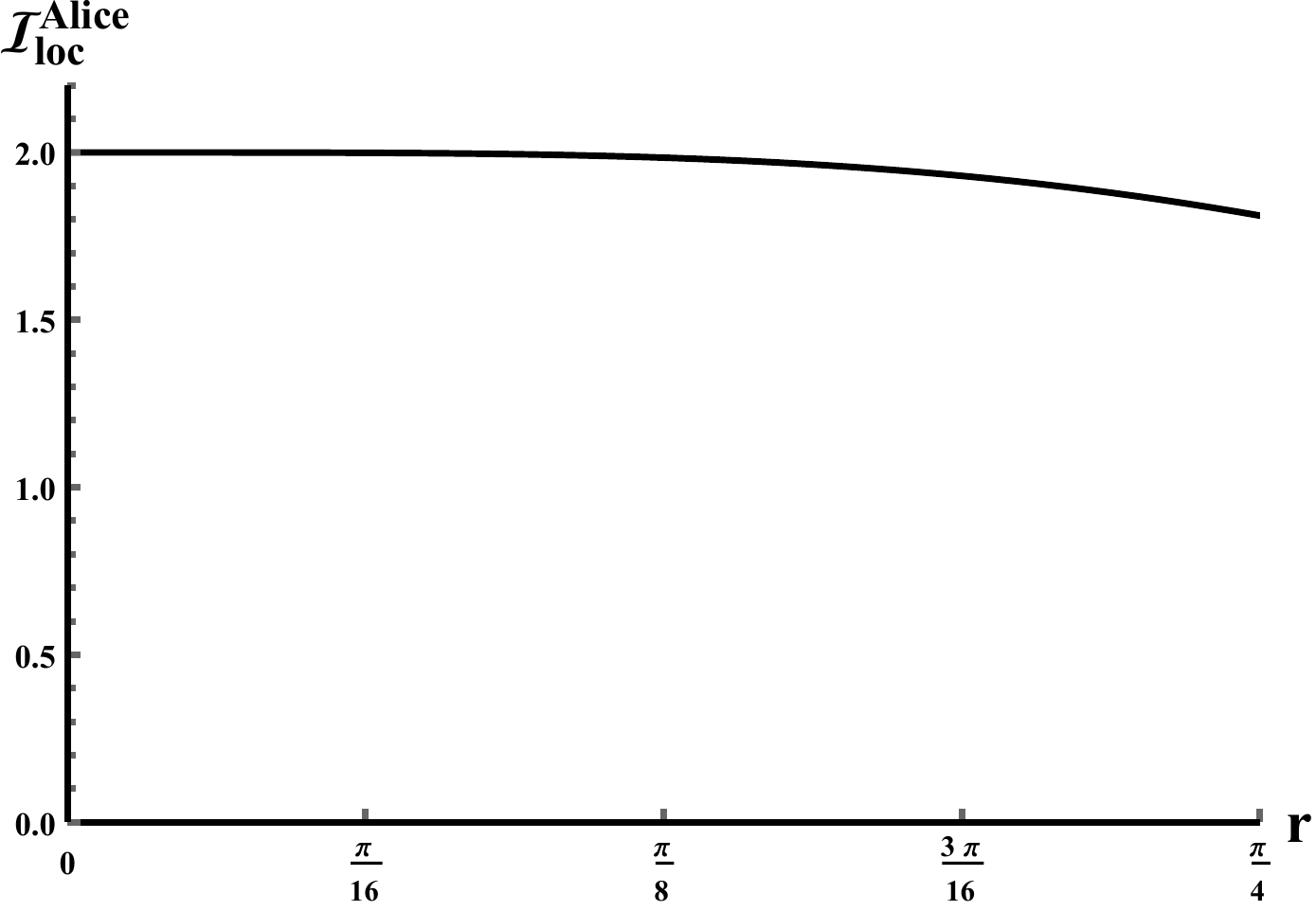}
		\label{FGWS1fig:locbemask}
	\end{subfigure}
	\caption{ The local and non-local information for the state \eqref{FGWS1rhoaccelerated}, where the  system is initially prepared in  Werner state with $x = y=z=-0.85$, for the dashed curve,  $x = y=z=- 0.75$, for the dotted curve and the  $X$- state with $x = - 0.6$, $y = - 0.2$ and $z = - 0.5$ for the solid curve.}
	\label{FGWS1fig:figureloc}
\end{figure}


\subsection{Only Alice's qubit is masked}
\label{Alice-accelerated-masked}
In this subsection we discuss the possibility of masking the information of Alice's qubit, by using the accelerated state (\ref{FGWS1rhoaccelerated}) where The final states  after Alice applies the masking process on her qubit is given by,

\begin{align}
\rho^{mask_a}_{\text{ac}}&=\frac{1}{4}\left\lbrace\right.
\left( \ket{000} \bra{000} + \ket{110} \bra{110} + \ket{111}\bra{111}\right)
\\&\quad\quad+
x \cos r \left( \ket{001}\bra{000} + \ket{000}\bra{001} - \ket{111}\bra{110} - \ket{110}\bra{111} \right)
\\&\quad\quad+
\frac{1}{2} ((z+1) \cos (2 r)+z-1) \left(  \ket{110}\bra{000} + \ket{000}\bra{110} \right)
\\&\quad\quad+
y \cos r \left( \ket{111}\bra{000} - \ket{110}\bra{001} - \ket{001}\bra{110} + \ket{000}\bra{111} +  \ket{001}\bra{001} \right) +
\\&\quad\quad
\frac{1}{2} (-(z-1) \cos (2 r)-z-1)  \left( \ket{111}\bra{001} +\ket{001}\bra{111} \right)
\left.\right\} .
\end{align}
Now by  tracing the third qubit of the state $\rho^{mask_a}_{123}$ one obtains,
\begin{align}
\rho^{mask_a}_{12}&=\frac{1}{2}\left( \ket{00} \bra{00} + \ket{11} \bra{11} \right) -\frac{1}{2}\sin^2 r \left(  \ket{11}\bra{00} + \ket{00}\bra{11} \right).
\label{FGWS1rho12}
\end{align}

The state \eqref{FGWS1rho12} is the same as those displayed in (12) and (16). Therefore it satisfies the  masking conditions, namely $\rho^{mask_a}_{1}=\tr_2\rho^{mask_a}_{12}=\frac{1}{2} I_{2}$ and $\rho^{mask_a}_{2}=\tr_1\rho^{mask_a}_{12}=\frac{1}{2} I_{2}$. Therefore, the masked information of Alice is encoded on the quantum correlation of the state\eqref{FGWS1rho12}. Moreover, the behavior of the quantum Fisher information,  the  local and  non-local information  is similar to that shown in Subsection (\ref{Both-accelereted-masked}).

The second and the third partitions are given by,
\begin{align}
\rho^{mask_a}_{13}&=\rho^{mask_a}_{23} \\
&=\frac{1}{4} \left( \ket{00}\bra{00}+\ket{10}\bra{10}+ \ket{01}\bra{01} + \ket{11}\bra{11}\right)
\\& \quad +	\frac{1}{4} (x \cos  r)\left( \ket{00}\bra{01} + \ket{01}\bra{00} - \ket{10}\bra{11} - \ket{11}\bra{10} \right) .
\label{FGWS1rho13}
\end{align}

It is clear that,  the state  \eqref{FGWS1rho13} satisfies the masking conditions, where $\rho^{mask_a}_{1}=\rho^{mask_a}_{3}=\frac{1}{2}I_{2}$. This means that, the masked information encoded on the classical/quantum correlation of the state $\rho^{mask_a}_{13}$. On the other hand,  the partition  \eqref{FGWS1rho13} represents  a separable state where the eigenvalues of $\rho^{T_{2}-mask_a}_{13} >0$.  Therefore, all the information are masked in the classical non-local state between Alice and Bob. Consequently, we evaluate the classical Fisher information which can be used to estimate the acceleration $r$ and the settings parameter $x$. For this aim one needs to evaluate the eigenvalues  of  \eqref{FGWS1rho13} which are given by,
\begin{align*}
&\lambda_{1}=\lambda_{2}=\frac{1}{4} (1-x \cos (r)),~ \lambda_{3}=\lambda_{4}=\frac{1}{4} (1+x \cos (r)). \\
\end{align*}
 The behavior of the classical Fisher information (CFI) with  respect to the parameter $r$ and $x$  can be evaluated analytically as
 \begin{eqnarray}
 \mathcal{F}_{C}^{r}&=&\frac{x^2\sin^2 r}{1-x^2\cos^2 r}, ~\quad
 \mathcal{F}_{C}^{x}=\frac{\cos^2r}{1-x^2\cos^2 r}.
 \label{FIFXrho13}
 \end{eqnarray}

 From these two forms it is clear that, $\mathcal{F}_{C}^{x}$ is undefined at either $x=\pm 1$ or $r=0$. Therefore it is impossible to estimate the initial state settings or the acceleration parameter when the initial accelerated system is prepared in a maximum entangled state. When $x=\pm 1$, the maximum value $\mathcal{F}_{C}^{r}=1$ is obtained while $\mathcal{F}_{C}^{x}=\cot^2r$, which tends to infinity at $r \rightarrow 0$.  Therefore,  the non-local product state \eqref{FGWS1rho13} can be used to estimate a class of  initial states which are defined by $|x|<1$ and arbitrary $y, z$. The behavior of the classical Fisher information $\mathcal{F}_{C}^{r}$ and $\mathcal{F}_{C}^{x}$  are displayed in   \figurename{(\ref{FGWS1fig:figure13maskr1})}. However, as shown in \figurename{(\ref{FGWS1fig:QFaftermaskingr})}, $\mathcal{F}_{C}^{r}$ increases as $r$ increases. The increasing rate depends on the  amount of entanglement contained in the initial accelerated state. As we mentioned above, analytically  the maximum value of $\mathcal{F}_{C}^{r}$ does not exceed one. The behavior  of $\mathcal{F}_{C}^{x}$ is shown in \figurename{(\ref{FGWS1fig:QFaftermaskingx})}, where it decreases gradually as the acceleration increases. The maximum values of $\mathcal{F}_{C}^{x}$ is displayed at $r=0$ and large initial entanglement, as clarified from the analytical form \eqref{FIFXrho13}.
The increasing and decreasing behavior of the classical Fisher information $\mathcal{F}_{C}^{r}$ and $\mathcal{F}_{C}^{x}$ is due to the entanglement. Since, the entanglement decreases as one increases the acceleration $r$, the classical correlation increases on the expanse of the quantum correlation. On the other hand the state \eqref{FGWS1rho13} contains only classical information, Therefore the CFI increases  with respect to the acceleration $r$.

{\normalsize
\begin{figure}[t!]
	\begin{subfigure}[t]{0.5\textwidth}
		\centering
		\caption{}
		\includegraphics[width=3in,height=3in]{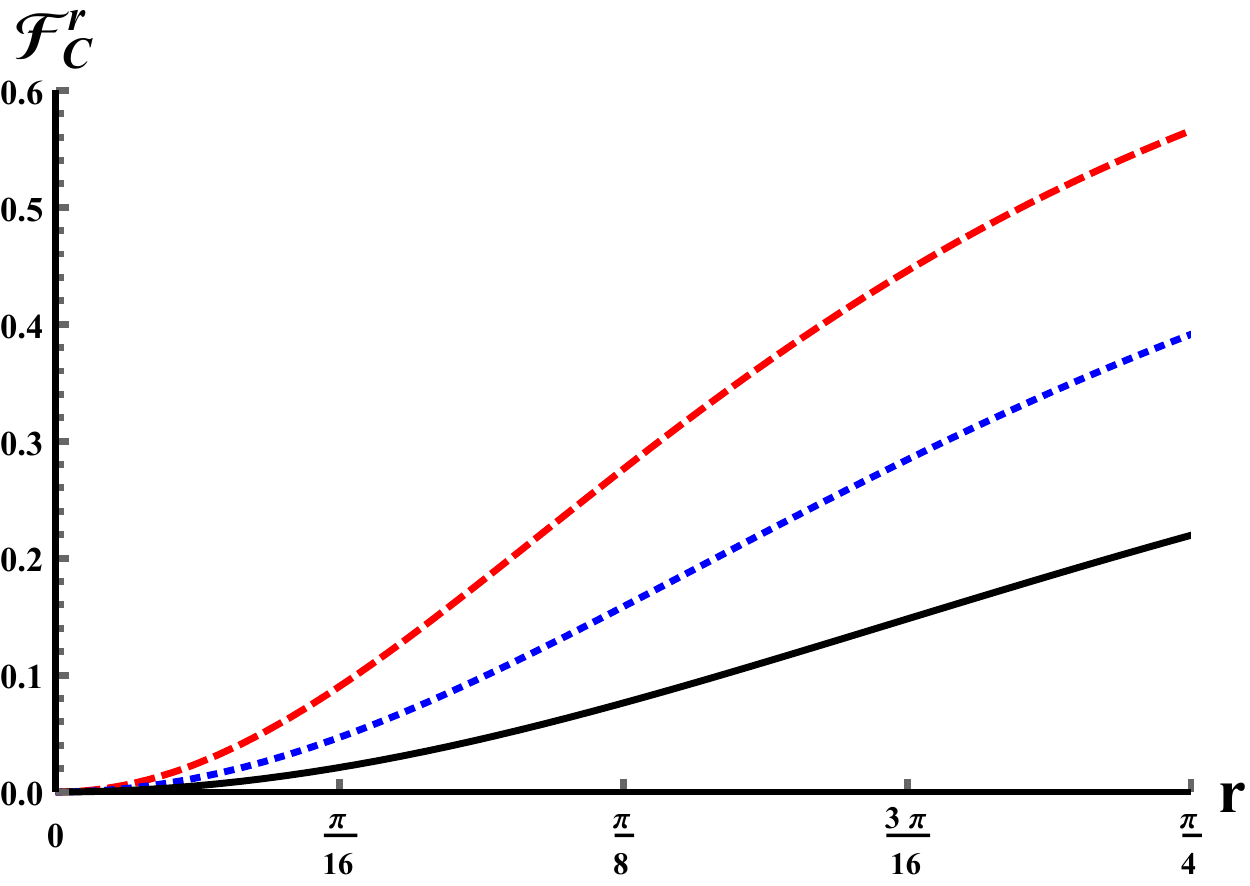}
		\label{FGWS1fig:QFaftermaskingr}
	\end{subfigure}
	~
	\begin{subfigure}[t]{0.5\textwidth}
		\centering
		\caption{}
		\includegraphics[width=3in,height=3in]{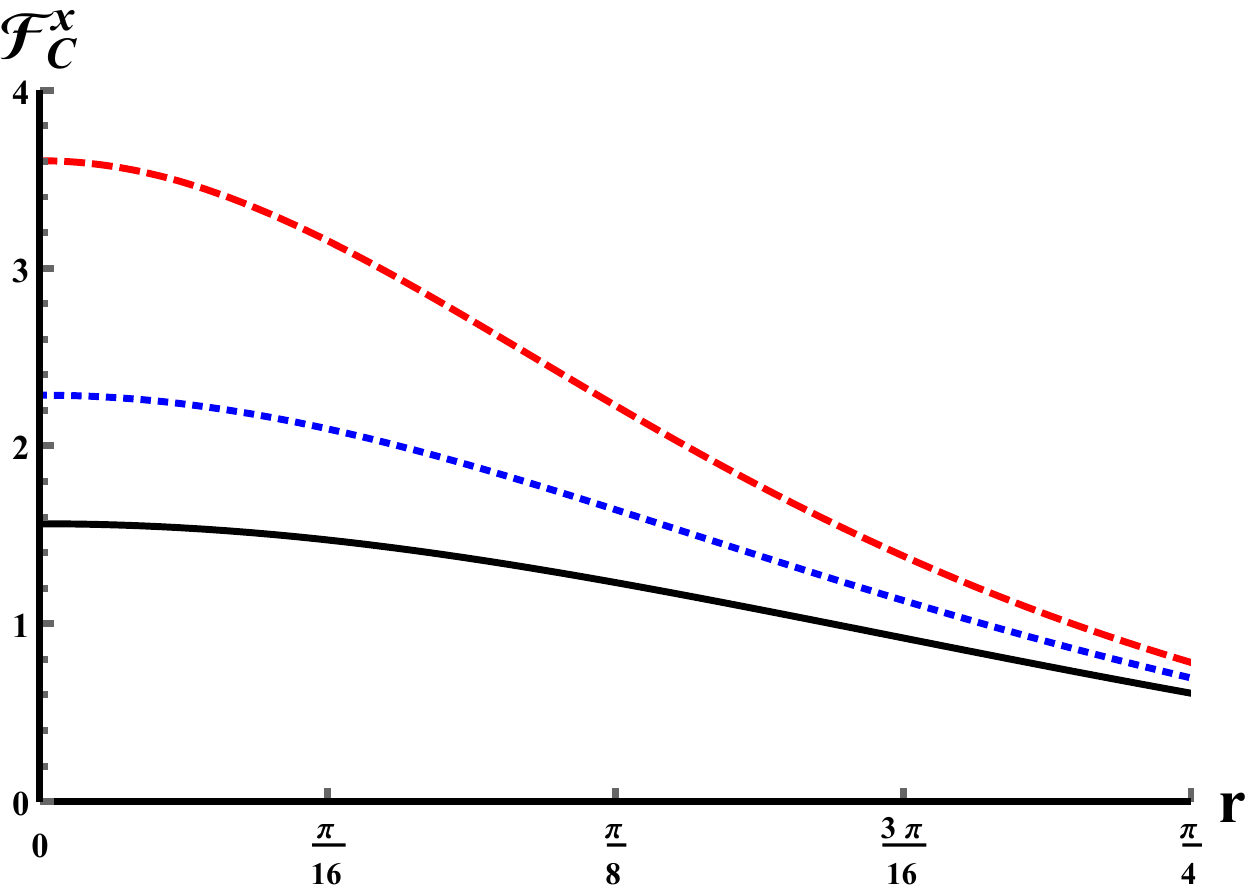}
		\label{FGWS1fig:QFaftermaskingx}
	\end{subfigure}
	\caption{ Estimating  (\subref{FGWS1fig:QFaftermaskingr}) the acceleration parameter $r$, and (\subref{FGWS1fig:QFaftermaskingx}) the $x$ parameter by using  \eqref{FGWS1rho13}. It is assumed that the system is initially prepared in a Werner state  with $x = y = z = -0.85$ for the dashed curve, $x = y = z = -0.75$ for the dotted curve and the $X$- state with $x = - 0.6$, $y = - 0.2$ and $z = - 0.5$ for the solid curve.}
	\label{FGWS1fig:figure13maskr1}
\end{figure}

}

The local and non-local information for \eqref{FGWS1rho13}  are shown in \figurename{(\ref{FGWS1fig13:figureloc})}. As it is displayed from \figurename{(\ref{FGWS1fig13:nlocbefmask})}, $\mathcal{I}_{nonloc}$  increases very slightly as $r$ increases, where the predicted amount of the non-local information is larger than that displayed in \figurename{(\ref{FGWS1fig:figureloc})}, namely before masking process which  means  it contains an extra information. On the other hand, there is no information encoded in the marginal states. These results are confirmed from the behavior of  the $\mathcal{I}_{loc}$ in \figurename{(\ref{FGWS1fig13:locbemask})}.
{\normalsize
\begin{figure}[t!]
	\begin{subfigure}[t]{0.5\textwidth}
		\centering
		\caption{}
		\includegraphics[width=3in,height=3in]{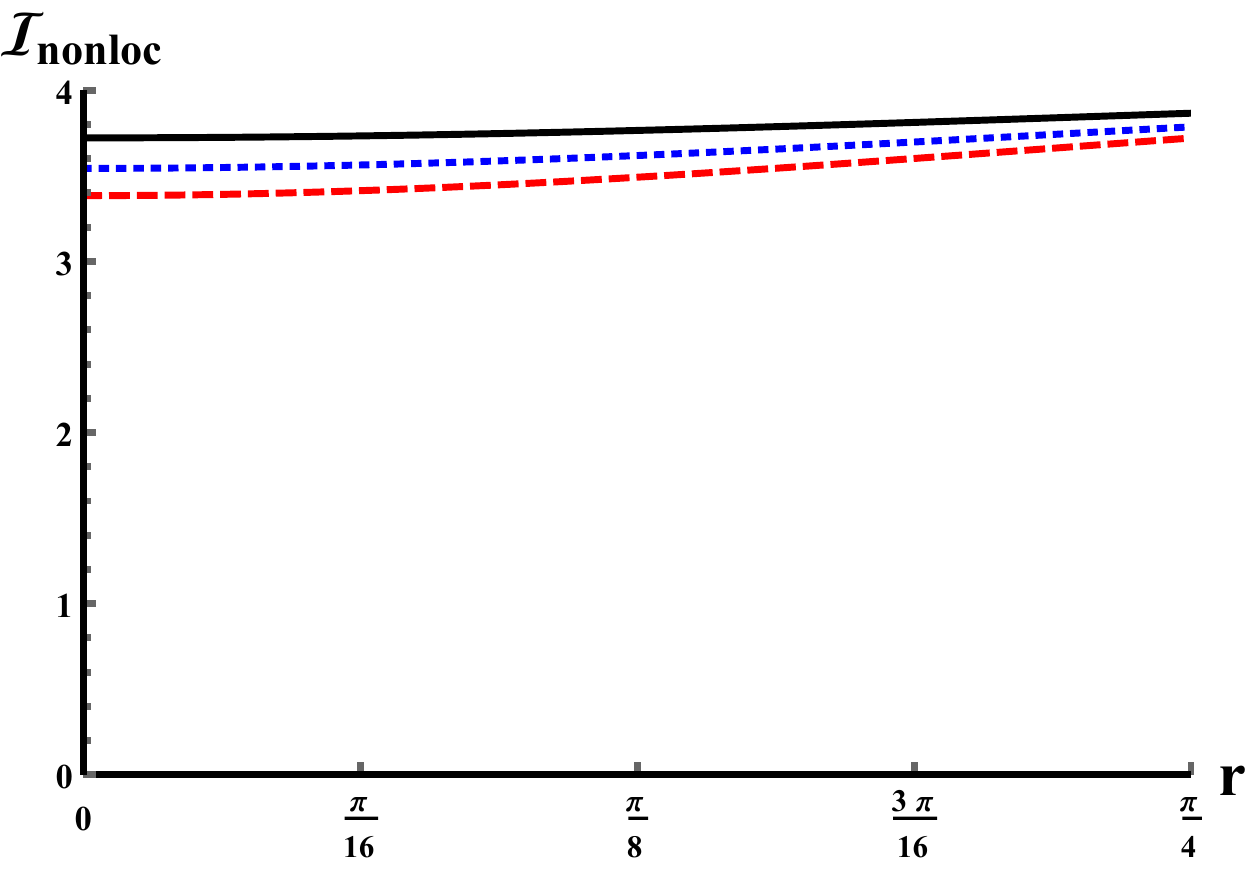}
		\label{FGWS1fig13:nlocbefmask}
	\end{subfigure}
	~
	\begin{subfigure}[t]{0.5\textwidth}
		\centering
		\caption{}
		\includegraphics[width=3in,height=3in]{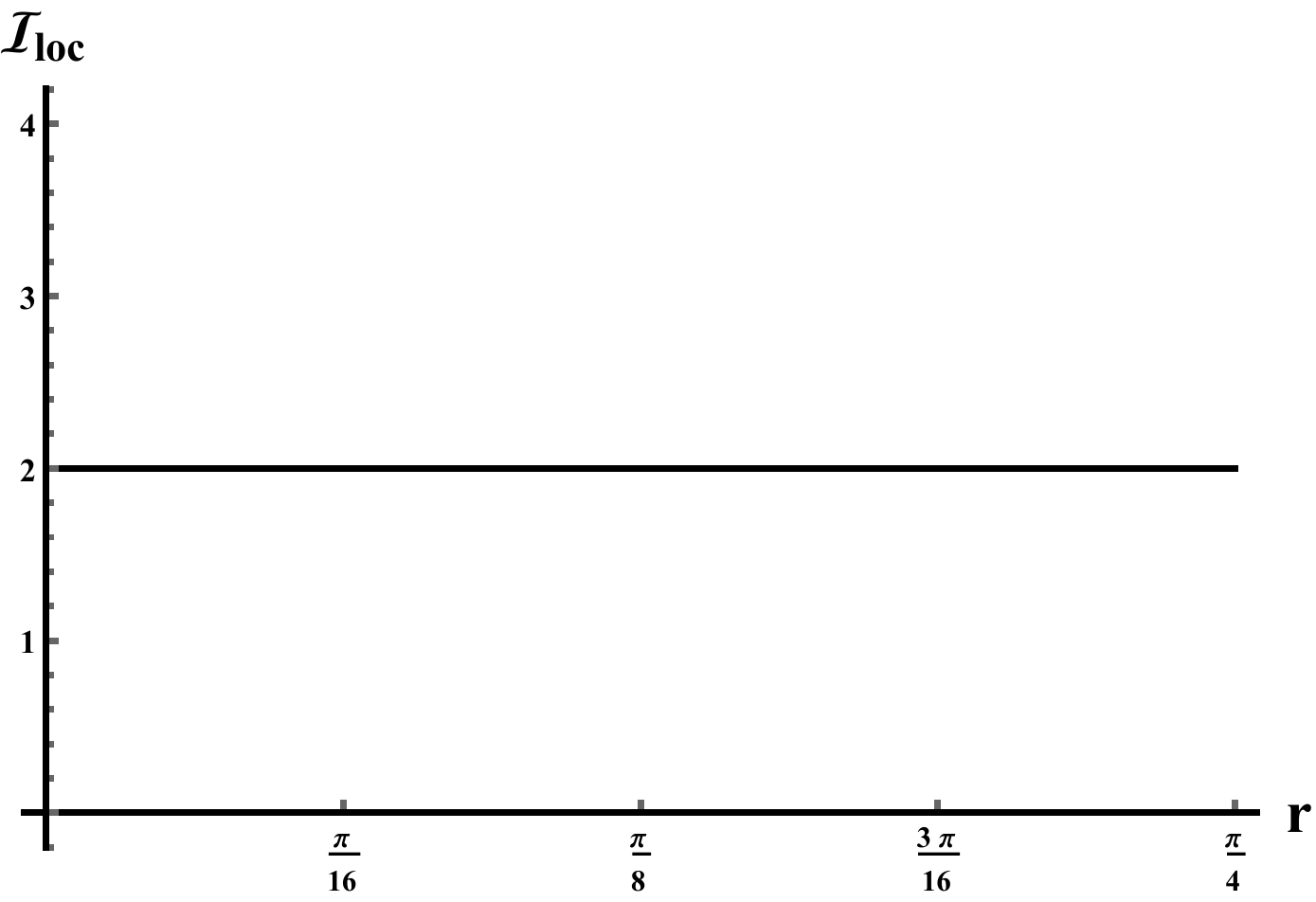}
		\label{FGWS1fig13:locbemask}
	\end{subfigure}
	\caption{The local and non-local information for \eqref{FGWS1rho13}}
	\label{FGWS1fig13:figureloc}
\end{figure}
}


\section{Accuracy of the masking process}
\label{fidelity}
\begin{figure}[t!]
	\begin{subfigure}[t]{0.3\textwidth}
		\centering
		\caption{}
		\includegraphics[width=2in,height=2in]{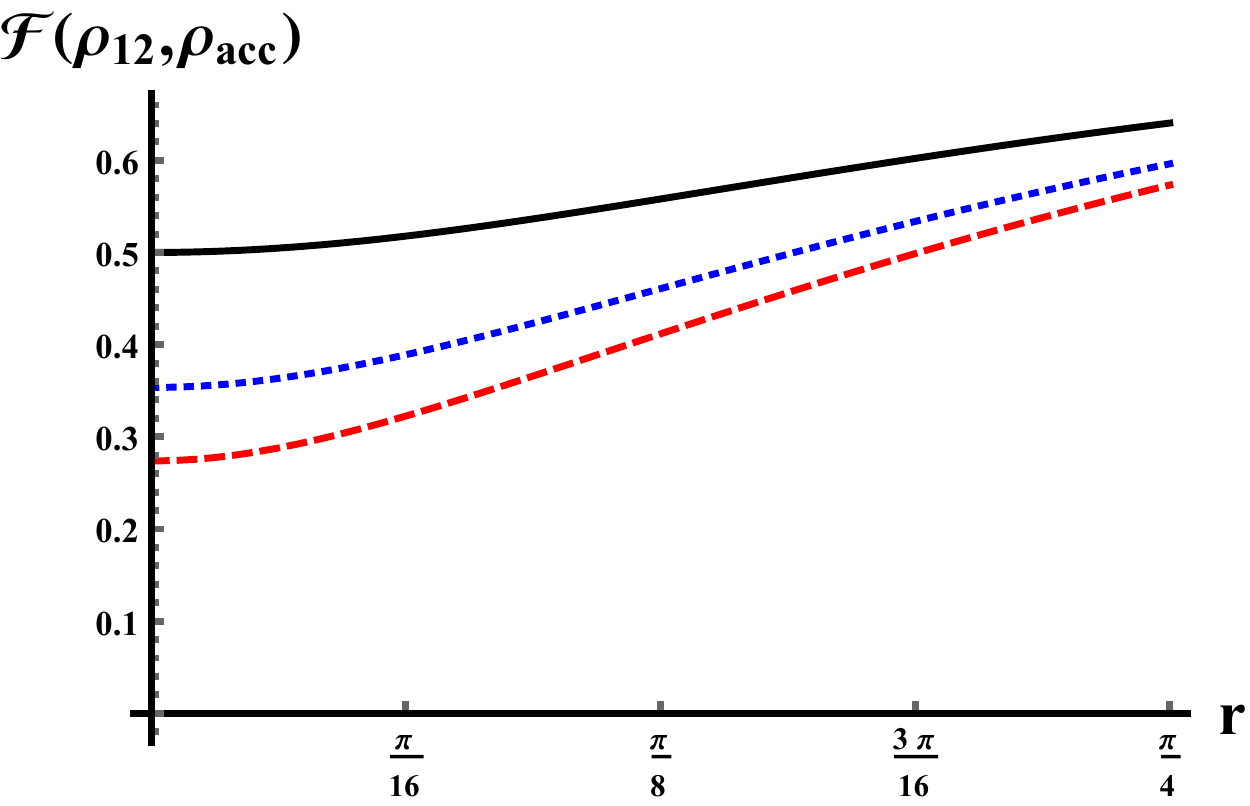}
		\label{FGWS9Fidelity12}
	\end{subfigure}
	~
	\begin{subfigure}[t]{0.3\textwidth}
		\centering
		\caption{}
		\includegraphics[width=2in,height=2in]{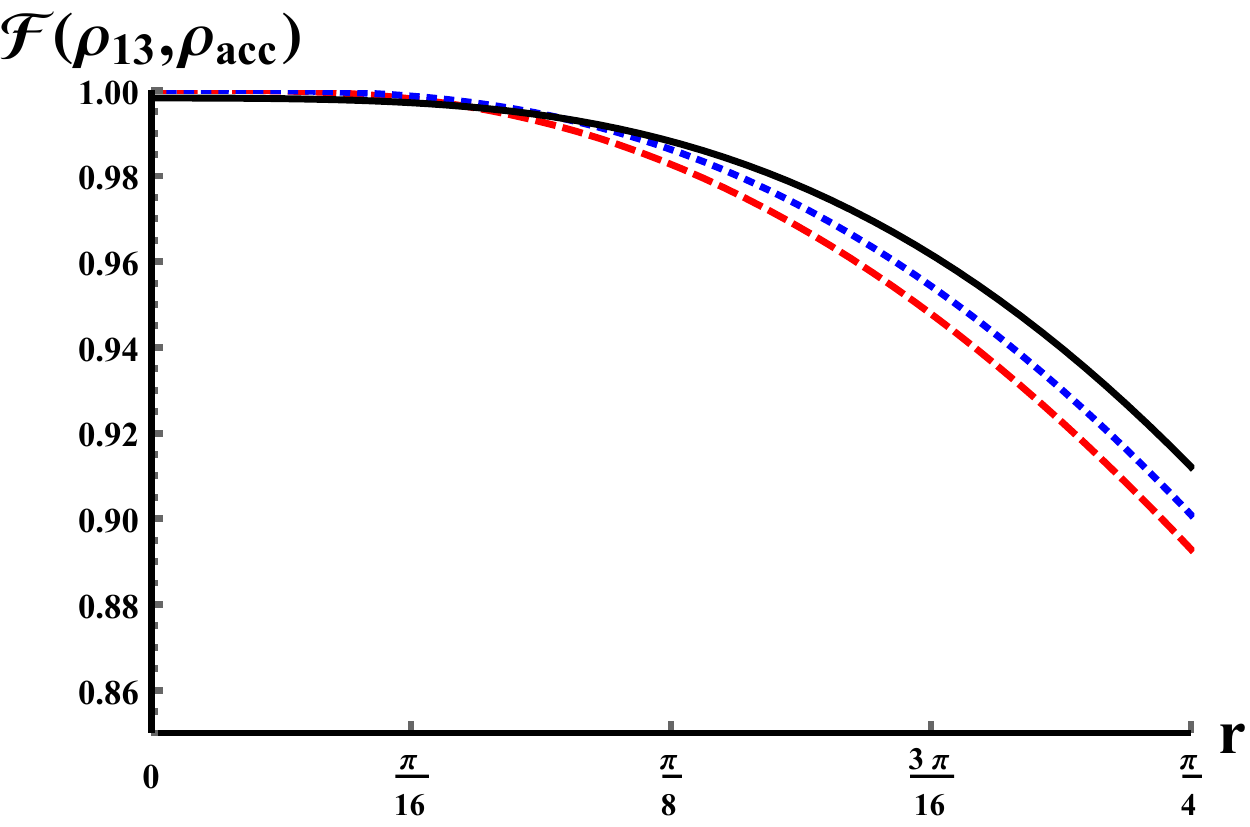}
		\label{FGWS9Fidelity13}
	\end{subfigure}
    ~
    \begin{subfigure}[t]{0.3\textwidth}
    	\centering
    	\caption{}
    	\includegraphics[width=2in,height=2in]{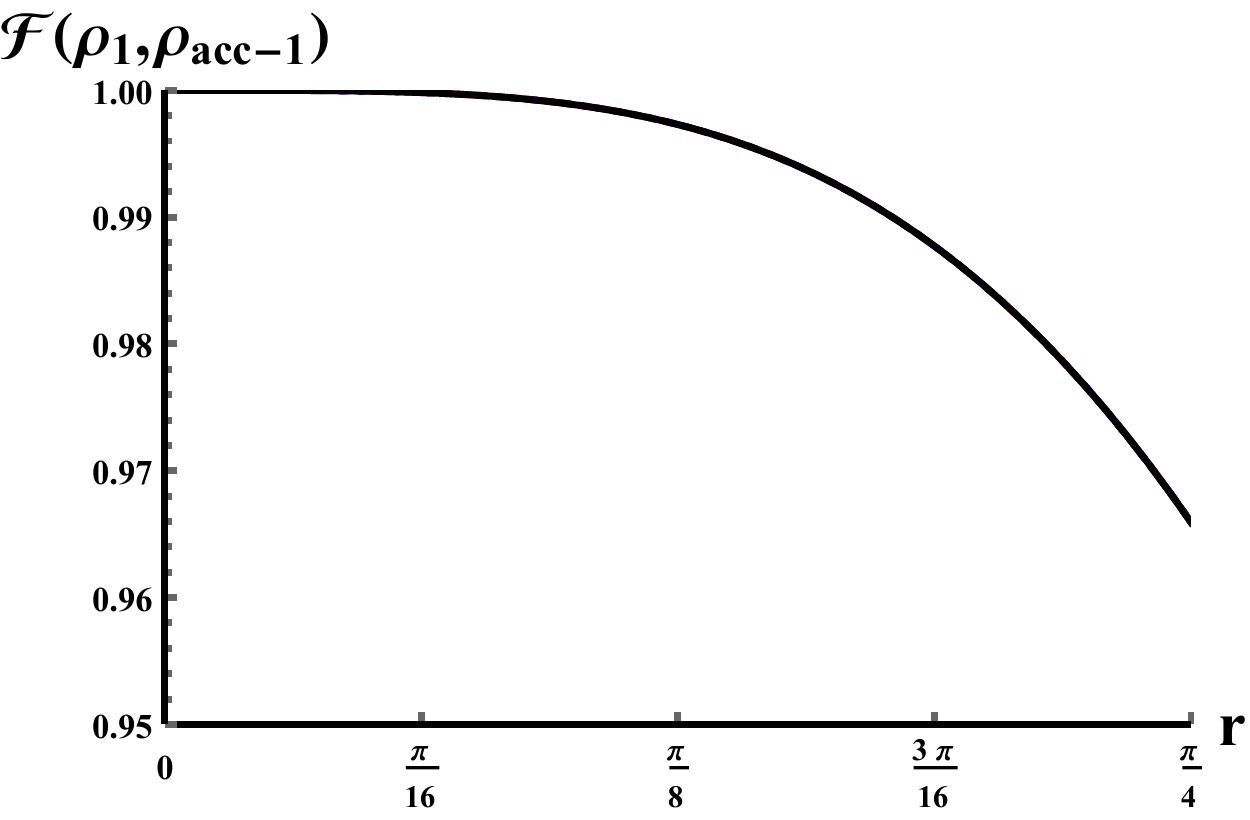}
    	\label{FGWS9Fidelity1}
    \end{subfigure}
	\caption{ The fidelity of the masked state by means of the generated (\subref{FGWS9Fidelity12}) local entangled state \eqref{FGWS9rho12}, (\subref{FGWS9Fidelity13}) the non-local separable state \eqref{FGWS9rho13} and (\subref{FGWS9Fidelity1}) the marginal states $\rho_{1(2)}$}
	\label{FGWS9Fidelity}
\end{figure}

It is worth to discuss the accuracy degree of the masking process by evaluating the fidelity of the masked accelerated state and its marginal.
Let us consider first case, where both qubits are accelerated and masked as in Subsection (\ref{Both-accelereted-masked}).  In this case, the generated masked state either, the entangled local state \eqref{FGWS9rho12} or the non-local separable state \eqref{FGWS9rho13}.  The fidelity of the masked state is displayed in \figurename{(\ref{FGWS9Fidelity})} by using the masked states \eqref{FGWS9rho12}  and \eqref{FGWS9rho13}. However, the fidelity evaluated by using the local entangled state \eqref{FGWS9rho12}, increases gradually as  the initial acceleration increases. Also, as it is displayed from \figurename{(\ref{FGWS9Fidelity12})}, the less entangled sate predicted large fidelity of the masked state. In \figurename{(\ref{FGWS9Fidelity13})}, we investigate the behavior of the fidelity of the masked state by using the non-local separable state for different initial accelerated state settings.  In general, the  behavior  shows that  the fidelity decreases as the initial increasing increases.  The smallest upper bounds of the fidelity is displayed for accelerated states having large initial entanglement.  The most important significant result that could be noticed from the behavior of the fidelity  is that: the fidelity  of the masked  state is maximum  at  small values of acceleration $r<\pi/8$.\\
In \figurename{(\ref{FGWS9Fidelity1})}, we plot the fidelity $\mathcal{F}_{\rho_{1(2)}}$ of the masked state of the marginal state $\rho_1$ or $\rho_2$. It is clear that the fidelity at $r<\pi/8$ is maximum, while at further values of $r$ the fidelity $\mathcal{F}_{\rho_{1(2)}}$ decreases quickly to reach its minimum value at $r=\pi/4$, where $min\{\mathcal{F}_{\rho_{1(2)}}\}>0.96$.\\

\begin{figure}[H]
	\begin{subfigure}[t]{0.5\textwidth}
		\centering
		\caption{}
		\includegraphics[width=2.5in,height=2.5in]{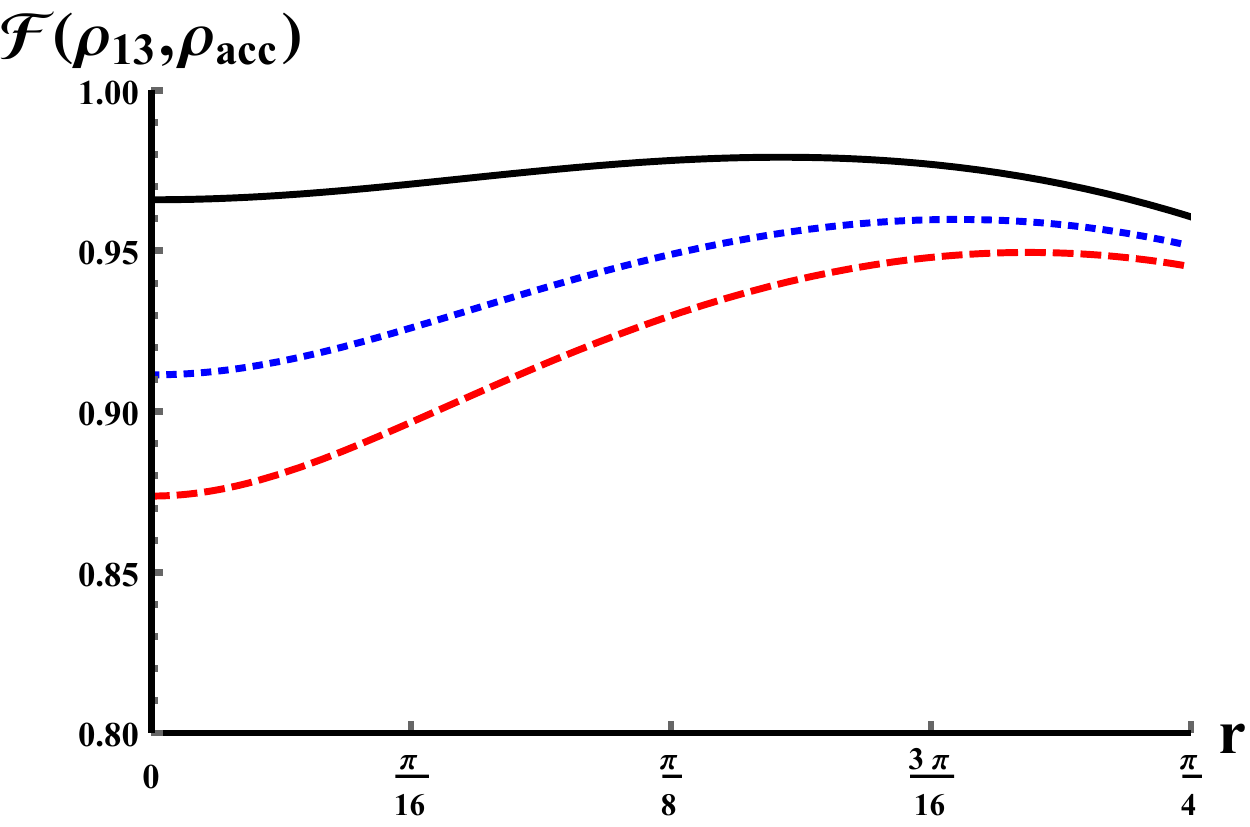}
		\label{FGWS7Fidelity13}
	\end{subfigure}
	~
	\begin{subfigure}[t]{0.5\textwidth}
		\centering
		\caption{}
		\includegraphics[width=2.5in,height=2.5in]{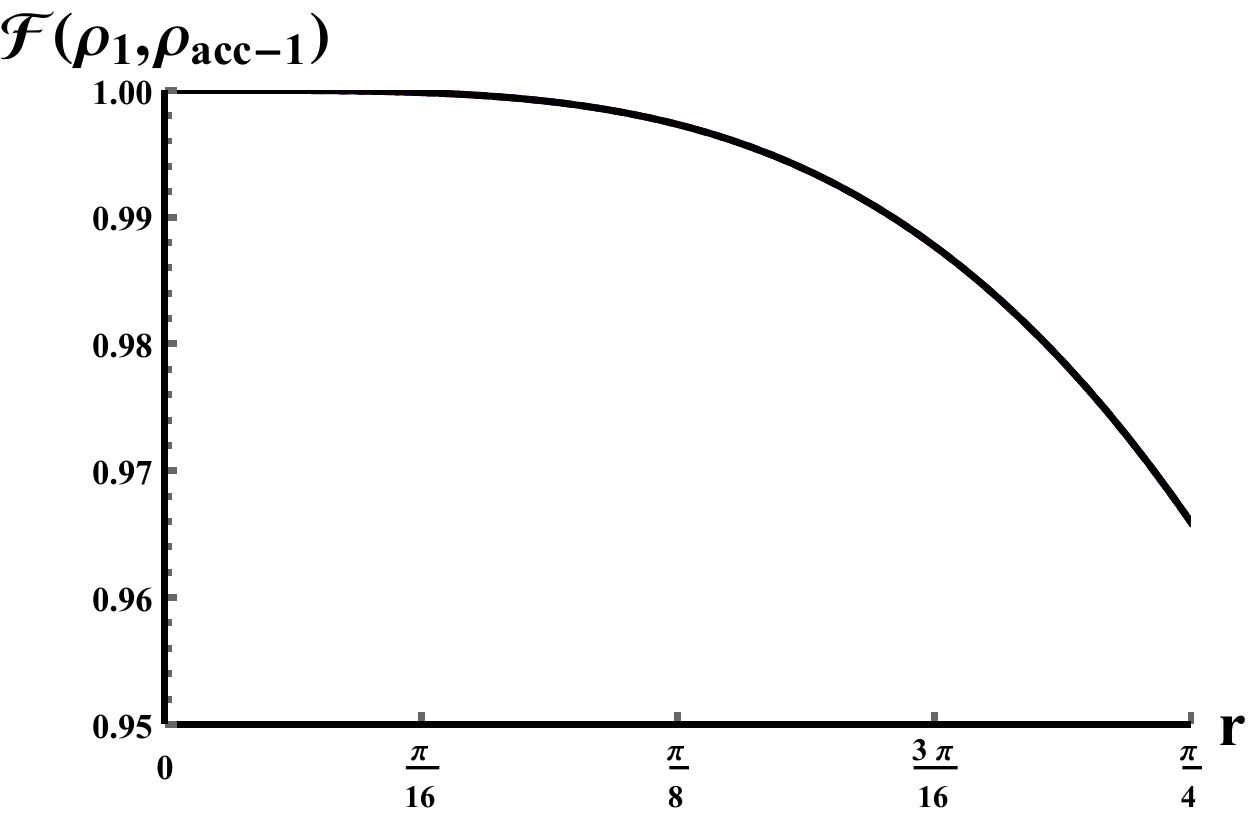}
		\label{FGWS7Fidelity1}
	\end{subfigure}
	\caption{ The fidelity of the masked state by means of the generated (\subref{FGWS7Fidelity13}) local entangled state \eqref{FGWS7rho13},   and (\subref{FGWS7Fidelity1}) the marginal states $\rho_{1(2)}$}
	\label{FGWS7Fidelity}
\end{figure}

Now, we assume that only Alice qubit is masked and both qubits are accelerated as studied in Subsection (\ref{Both-accelereted-Alice-masked}). In this case, only the non-local partition is different. So, it is enough to consider the fidelity of the accelerated state by means of the masked state  \eqref{FGWS7rho13}. As it is displayed from \figurename{(\ref{FGWS7Fidelity13})}, the fidelity increases as the acceleration parameter $r$ increases. Similarly, the possibility of masking the less entangled state is better than that displayed for those with large  initial entanglement. The minimum fidelity is predicted when the acceleration tends to infinity  is larger than $91\%$. The fidelity of the masked marginal state of Alice is displayed in \figurename{(\ref{FGWS7Fidelity1})}. The general behavior is similar to that displayed in \figurename{(\ref{FGWS9Fidelity})}, where the minimum fidelity is larger than $96.05\%$ and it is maximum, i.e. $(=1)$ at any $r<\pi/8$.

Also, our third case, namely: only Alice's qubit is accelerated and masked as in Subsection (\ref{Alice-accelerated-masked});  predicates the same behavior of the fidelity. From our previous results, one may conclude that one can mask the accelerated information  of two qubit system with high efficiency via the non-local separable state. The minimum masking efficiency of the marginal  state exceeds $96\%$, if both qubits or a single qubit is accelerated. However, for the total state the minimum  fidelity of the masked state masking  is larger than $90\%$ if both qubits are accelerated and above $85.5\%$  when only one qubit is accelerated. Therefore the high  efficiency of the masking process is achieved, when both subsystems are accelerated and masked.

\section{Conclusion}
\label{conclusion}
In this contribution, we discussed  the possibility of applying the simplest masking protocol on accelerated  two-qubit system. In this treatment,  different scenarios  are considered: either the two qubits are accelerated or only one of them, where the masking process is performed on the accelerated qubit only.  To clarify this idea we assume that, the accelerated system is prepared in different settings of the $X$-state.  The  amount of the non classical correlations of the accelerated systems  are quantified by means of the negativity measure. Moreover, the quantum Fisher information is used to  estimate the parameters which describe the accelerated two-qubit system. Further, we quantify the accelerated local and non-local information which are encoded in the accelerated system.

Due to the masking process on the accelerated subsystems, there are  different  partitions of the masked accelerated systems which are generated. It is shown that, there  is an entangled  state which is  generated locally  between each accelerated qubit and the  masker qubit. There are  different types of classical correlated states which are generated non-locally. However, all these kinds of states local entangled/ non-local separable, satisfy the masking conditions.  It is explored that, the local information of  the accelerated state   is encoded in the local entangled state and the non-local product states.

As a result of the masking process, the local entangled state depends only on the acceleration parameter, while the non-local separable state depends on the acceleration parameter and a single parameter of the initial state settings.  Therefore, the possibility of masking a large  set of initial  types of the $X$-states is one of the most important outcomes of the masking process. The estimation process of the acceleration parameter may be achieved locally via the entangled state or non-locally by using the non-local separable state. However, for the initial state settings parameter it can be estimated only by using the non-local  separable state

The behavior of the local and non-local information is estimated before  and after the masking process. It is shown that, due to this process the local information can not be  quantified  locally. This means that, the local information is masked in the local-entangled state, and consequently the total information that encoded in this state includes the local and non-local information.  Similarly, the non-local product states can be used to quantify the local and the non-local information. These results explore an alternative technique to hide the information from the eavesdroppers, where to get any information Eve has to reach the non-local product state, which is impossible, where each part is in a different lab.

The fidelity of the masked state is evaluated for all the suggested cases. It is shown that, the fidelity of the local entangled  masked state increases as the acceleration increases, while that for the non-local separable states and their marginal subsystems decreases gradually. The  fidelity  predicted for the accelerated  masked state is  maximum at small values of the initial acceleration. The minimum fidelity  displayed by means of the non-local separable states and their marginal is more than $96\%$.

In this context, it is worth to mention that the generated masked state displays an unexpected different behavior. It is well known that, the acceleration process has a decoherence effect. However, the amount of entanglement of the masked entangled state increases as the acceleration increases.  Moreover, the estimation degree of the acceleration parameter increases as the initial acceleration increases.

In conclusion: it is possible to mask the accelerated information encoded in a two-qubit system by using a minimum number of masker qubits. The generated masked state may be locally entangled/ separable state, where these states satisfy  the masking and hiding information, respectively.
However, the fidelity of masking the accelerated whole states and its subsystems locally is above $96\%$, while it is maximum for small values of acceleration.  We expect that these results may be important in the context of storing and handling the quantum information securely.


\begin{thebibliography}{10}
	
	\bibitem{Gisin2002}
	N.~Gisin, G.~Ribordy, W.~Tittel, and H.~Zbinden.
	\newblock {Quantum cryptography}.
	\newblock {\em Reviews of Modern Physics}, 74(1):145--195, 2002.
	
	\bibitem{Fink2017}
	M.~Fink, A~Rodriguez, J.~Handsteiner, A.~Ziarkash, F.~Steinlechner, T.~Scheidl,
	I.~Fuentes, J.~Pienaar, T.~C. Ralph, and R.~Ursin.
	\newblock {Experimental test of photonic entanglement in accelerated reference
		frames}.
	\newblock {\em Nature Communications}, 8(1):1--6, 2017.
	
	\bibitem{Schumacher1995}
	B.~Schumacher.
	\newblock {Quantum coding}.
	\newblock {\em Physical Review A}, 51(4):2738--2747, 1995.
	
	\bibitem{Modi2018}
	K.~Modi, A.~K. Pati, A.~Sen, and U.~Sen.
	\newblock {Masking quantum information is impossible}.
	\newblock {\em Physical Review Letters}, 120(23), 2018.
	
	\bibitem{Ghosh2019}
	T.~Ghosh, S.~Sarkar, B.~K. Behera, and P.~K. Panigrahi.
	\newblock {Masking of quantum information is possible}.
	\newblock {\em arXiv preprint arXiv:1910.00938}, 2019.
	
	\bibitem{Li2019}
	M.~S. Li and Y.~L. Wang.
	\newblock {Masking quantum information in multipartite scenario}.
	\newblock {\em Physical Review A}, 98(6), 2018.
	
	\bibitem{Liang2020}
	X.~B. Liang, B.~Li, S.~M. Fei, and H.~Fan.
	\newblock {Impossibility of masking a set of quantum states of nonzero
		measure}.
	\newblock {\em Physical Review A}, 101(4):042321, 2020.
	
	\bibitem{Liang2019}
	X.~B. Liang, B.~Li, and S.~M. Fei.
	\newblock {Complete characterization of qubit masking}.
	\newblock {\em Physical Review A}, 100(3), 2019.
	
	\bibitem{Lie2019}
	S.~H. Lie, H.~Kwon, M.~S. Kim, and H.~Jeong.
	\newblock {Unconditionally secure qubit commitment scheme using quantum
		maskers}.
	\newblock {\em arXiv preprint arXiv:1903.12304}, 2019.
	
	\bibitem{Fuentes-Schuller2005}
	I.~Fuentes and R.~B. Mann.
	\newblock {Alice falls into a black hole: Entanglement in noninertial frames}.
	\newblock {\em Physical Review Letters}, 95(12), 2005.
	
	\bibitem{Alsing2006}
	P.~M. Alsing, I.~Fuentes, R.~B. Mann, and T.~E. Tessier.
	\newblock {Entanglement of dirac fields in noninertial frames}.
	\newblock {\em Physical Review A - Atomic, Molecular, and Optical Physics},
	74(3):1--15, 2006.
	
	\bibitem{Moradi2009}
	S.~Moradi.
	\newblock {Distillability of entanglement in accelerated frames}.
	\newblock {\em Physical Review A - Atomic, Molecular, and Optical Physics},
	79(6):064301, 2009.
	
	\bibitem{Landulfo2009}
	A.G.S. Landulfo and G.E.A. Matsas.
	\newblock {Sudden death of entanglement and teleportation fidelity loss via the
		Unruh effect}.
	\newblock {\em Physical Review A - Atomic, Molecular, and Optical Physics},
	80(3):032315, 2009.
	
	\bibitem{Martin-Martinez2010a}
	E.~Mart{\'{i}}n-Mart{\'{i}}nez and J.~Le{\'{o}}n.
	\newblock {Quantum correlations through event horizons: Fermionic versus
		bosonic entanglement}.
	\newblock {\em Physical Review A - Atomic, Molecular, and Optical Physics},
	81(3):032320, 2010.
	
	\bibitem{metwally2016}
	N.~Metwally.
	\newblock {Entanglement of simultaneous and non-simultaneous accelerated
		qubit-qutrit systems}.
	\newblock {\em Quantum Information and Computation}, 16(5-6):530--542, 2016.
	
	\bibitem{metwally2013}
	N.~Metwally.
	\newblock {Usefulness classes of traveling entangled channels in noninertial
		frames}.
	\newblock {\em International Journal of Modern Physics B}, 27(28), 2013.
	
	\bibitem{Hamdoun2020}
	H.~Hamdoun and A.~Sagheer.
	\newblock {Information security through controlled quantum teleportation
		networks}.
	\newblock {\em Digital Communications and Networks}, In press, jun 2020.
	
	\bibitem{metwally2014}
	N.~Metwally and A.~Sagheer.
	\newblock {Quantum coding in non-inertial frames}.
	\newblock {\em Quantum Information Processing}, 13(3):771--780, 2014.
	
	\bibitem{Bradler2008}
	K.~Bradler, P.~Hayden, and P.~Panangaden.
	\newblock {Private information via the Unruh effect}.
	\newblock {\em Journal of High Energy Physics}, 2009(8), 2008.
	
	\bibitem{Yu2007a}
	T.~Yu and J.~H. Eberly.
	\newblock {Evolution from entanglement to decoherence of bipartite mixed "X"
		states}.
	\newblock {\em Quantum Information and Computation}, 7(5-6):459--468, 2007.
	
	\bibitem{Martin-Martinez2011}
	E.~Mart{\'{i}}n-Mart{\'{i}}nez and I.~Fuentes.
	\newblock {Redistribution of particle and antiparticle entanglement in
		noninertial frames}.
	\newblock {\em Physical Review A - Atomic, Molecular, and Optical Physics},
	83(5), 2011.
	
	\bibitem{Peres1996}
	A.~Peres.
	\newblock {Separability criterion for density matrices}.
	\newblock {\em Physical Review Letters}, 77(8):1413--1415, 1996.
	
	\bibitem{Horodecki1996}
	R.~Horodecki, M.~Horodecki, and P.~Horodecki.
	\newblock {Teleportation, Bell's inequalities and inseparability}.
	\newblock {\em Physics Letters, Section A: General, Atomic and Solid State
		Physics}, 222(1-2):21--25, 1996.
	
	\bibitem{Yao2014}
	Y.~Yao, X.~Xiao, L.~Ge, X.~G. Wang, and C.~P. Sun.
	\newblock {Quantum fisher information in noninertial frames}.
	\newblock {\em Physical Review A - Atomic, Molecular, and Optical Physics},
	89(4):1--7, 2014.
	
	\bibitem{Metwally2016a}
	N.~Metwally.
	\newblock {Enhancing entanglement, local and non-local information of
		accelerated two-qubit and two-qutrit systems via weak-reverse measurements}.
	\newblock {\em Eur. Phys. Lett.}, 116(6), 2016.
	
	\bibitem{Abd-Rabbou2019a}
	M.~Y. Abd-Rabbou, N.~Metwally, M.~M.A. Ahmed, and A.~S.F. Obada.
	\newblock {Suppressing the information losses of accelerated qubit-qutrit
		system}.
	\newblock {\em International Journal of Quantum Information}, 17(4), 2019.
	
\end{thebibliography}
\end{document}